\documentclass[journal=jacsat,manuscript=article]{achemso}

\SectionNumbersOn 
\usepackage[version=3]{mhchem} 
 \usepackage[utf8]{inputenc}
 \usepackage[T1]{fontenc}
 \usepackage{lmodern}
\usepackage{amsmath}
\usepackage{amsfonts}
\usepackage{amssymb}
\usepackage{esvect}
\usepackage{graphicx}
\usepackage{textgreek}
\usepackage[position=top]{subfig}
\usepackage{pdfpages}
\usepackage{booktabs}
\usepackage{geometry}
\usepackage{mathtools}
\usepackage{colortbl}
\usepackage{pdflscape}
\usepackage{adjustbox}
\usepackage{lipsum}
\usepackage{colortbl}
\usepackage{xcolor}
\definecolor{myred}{RGB}{255, 153, 153} 
\definecolor{myblue}{RGB}{204, 204, 255}
\definecolor{mygreen}{RGB}{153, 255, 153}
\definecolor{mypurple}{RGB}{204, 153, 255}
\definecolor{mysilver}{RGB}{220,220,220}

\usepackage{cancel}
\usepackage{soul}

\newcommand{\bmu}{\boldsymbol{\mu}}

\newcommand{\bE}{\mathbf{E}}
\newcommand{\bT}{\mathbf{T}}

\newcommand\addtag{\refstepcounter{equation}\tag{\theequation}} 
\usepackage{xspace}
\newcommand*{\thead}[1]{\multicolumn{1}{c}{#1}}
\newcommand{\EctREG}[1]{\ensuremath{E^{(#1)}_{\rm cd}(\rm Reg)}\xspace}
\newcommand{\DELTAHF}[0]{\ensuremath{\rm{{\delta}_{HF}^{(2)}}}\xspace}
\newcommand{\Ssq}{\ensuremath{S^2}\xspace}

\newcommand{\almo}{{ALMO}\xspace}
\newcommand{\sapt}{{SAPT}\xspace}
\newcommand{\saptdft}{{SAPT(DFT)}\xspace}
\newcommand{\ccsdt}{{CCSD(T)}\xspace}

\newcommand{\sibfa}{{SIBFA21}\xspace}

\newcommand{\amoeba}{{AMOEBA14}\xspace}
\newcommand{\amoebap}{{AMOEBA+}\xspace}

\newcommand{\mbucb}{{MB-UCB-MDQ}\xspace}
\newcommand{\mbpol}{{MB-POL}\xspace}
\newcommand{\tinkerhp}{{TINKER-HP}\xspace}
\newcommand{\gem}{{GEM}\xspace}

\newcommand{\dens}{\ensuremath{\rho}\xspace}
\newcommand{\vap}{\ensuremath{\Delta {H}_{\rm vap}}\xspace}
\newcommand{\dielec}{\ensuremath{\epsilon_{0}}\xspace}
\newcommand{\isocomp}{\ensuremath{\kappa_{T}}\xspace}
\newcommand{\heat}{\ensuremath{C_p}\xspace}
\newcommand{\diff}{\ensuremath{D}\xspace}

\newcommand{\kcalmol}{kcal mol$^{-1}$\xspace}

\newcommand{\ecp}{cal mol$^{-1}$ K$^{-1}$\xspace}
\newcommand{\ediff}{10$^{-5}$ cm$^{2}$ s$^{-1}$\xspace}

\newcommand{\etal}{\emph{et al.}\xspace}
\newcommand{\abinitio}{{\em ab initio}\xspace}

\newcommand{\polFF}{{polFF}\xspace}
\newcommand{\polFFs}{{polFFs}\xspace}

\newcommand{\figrff}[1]{Figure~\ref{#1}}
\newcommand{\tabrff}[1]{Table~\ref{#1}}

\newcommand{\Eqrff}[1]{Eq.~\eqref{#1}}

\newcommand{\erep}{\ensuremath{E_{rep}}\xspace}
\newcommand{\epol}{\ensuremath{E_{pol}}\xspace}

\newcommand{\ect}{\ensuremath{E_{ct}}\xspace}
\newcommand{\edisp}{\ensuremath{E_{disp}}\xspace}
\newcommand{\ebind}{\ensuremath{E_{bind}}\xspace}

\author{Sehr Naseem-Khan}
\affiliation[SorbonneU]{Sorbonne Université, LCT, UMR 7616 CNRS, 75005, Paris, France}
\alsoaffiliation[UNT]{Department of Chemistry, University of North Texas, Denton, TX 76201, USA}
\author{Louis Lagardère}
\affiliation[SorbonneU]{Sorbonne Université, LCT, UMR 7616 CNRS, 75005, Paris, France}
\alsoaffiliation[IP2CT]{Sorbonne Université, IP2CT, FR 2622 CNRS,75005, Paris, France}
\email{louis.lagardere@sorbonne-universite.fr}
\author{Christophe Narth}
\affiliation[SorbonneU]{Sorbonne Université, LCT, UMR 7616 CNRS, 75005, Paris, France}
\author{\\ G. Andrés Cisneros}
\affiliation[UNT]{Department of Chemistry, University of North Texas, Denton, TX 76201, USA}
\alsoaffiliation[UTD]{Present adress: Department of Physics, University of Texas at Dallas, TX 75080, USA}
\author{Pengyu Ren}
\affiliation[UT]{Department of Biomedical Engineering, The University of Texas at Austin, TX 78712, USA}
\author{Nohad Gresh}
\affiliation[SorbonneU]{Sorbonne Université, LCT, UMR 7616 CNRS, 75005, Paris, France}
\email{nohad.gresh@lct.jussieu.fr}
\author{Jean-Philip Piquemal}
\affiliation[SorbonneU]
{Sorbonne Université, LCT, UMR 7616 CNRS, 75005, Paris, France}
\alsoaffiliation[IUF]{Institut Universitaire de France, 75005, Paris, France}
\alsoaffiliation[UT]{Department of Biomedical Engineering, The University of Texas at Austin, TX 78712, USA}
\email{jean-philip.piquemal@sorbonne-universite.fr}

\title[SIBFA]
  {Development of the Quantum Inspired SIBFA Many-Body Polarizable Force Field: Enabling Condensed Phase Molecular Dynamics Simulations}

\begin{document}

\begin{tocentry}




\centering
\begin{center}
\vfill
    \includegraphics[scale=0.2]{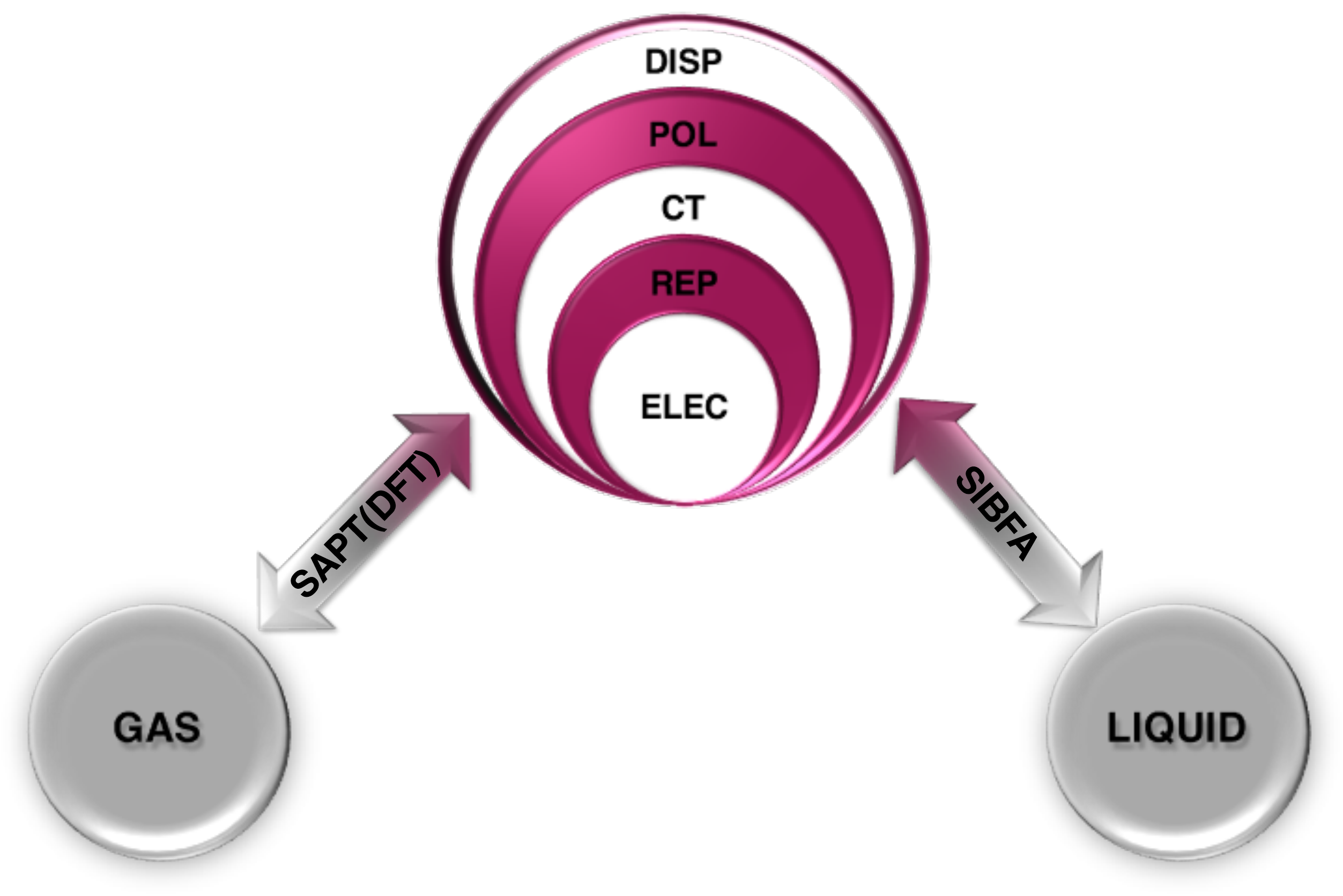} 
\end{center}

\end{tocentry}


%

\begin{abstract} 

We present the extension of the SIBFA (Sum of Interactions Between Fragments \textit{Ab initio} Computed) many-body polarizable force field to condensed phase Molecular Dynamics (MD) simulations. 
The Quantum-Inspired SIBFA procedure is grounded on simplified integrals obtained from localized molecular orbital theory and achieves full separability of its intermolecular potential. 
It embodies long-range multipolar electrostatics (up to quadrupole) coupled to a short-range penetration correction (up to charge-quadrupole), exchange-repulsion, many-body polarization, many-body charge transfer/delocalization, exchange-dispersion and dispersion (up to C$_{10}$). 
This enables the reproduction of all energy contributions of \abinitio Symmetry-Adapted Perturbation Theory (\saptdft) gas phase reference computations. The SIBFA approach has been integrated within the Tinker-HP massively parallel MD package. To do so all SIBFA energy gradients have been derived and the approach has been extended to enable periodic boundary conditions simulations using Smooth Particle Mesh Ewald. This novel implementation also notably includes a computationally tractable simplification of the many-body charge transfer/delocalization contribution. 
As a proof of concept, we perform a first computational experiment defining a water model fitted on a limited set of \saptdft data.
SIBFA is shown to enable a satisfactory reproduction of both gas phase energetic contributions and condensed phase properties highlighting the importance of its physically-motivated functional form.     
\end{abstract}

\section{Introduction}

A major incentive for the development of polarizable force fields (\polFFs) is their wide range of applications up to very large molecular assemblies encompassing highly charged ones.
\cite{stone2013theory,NohadGresh2007,warshelrev,doi:https://doi.org/10.1002/9781118889886.ch2,10.3389/fmolb.2019.00143,doi:10.1146/annurev-biophys-070317-033349, doi:10.1021/acs.chemrev.8b00763} 
However, as for any empirical approach, their accuracy and transferability should be duly evaluated first on the basis of reference \abinitio Quantum Chemistry (QC). 
The refinement of a polarizable water model is considered an essential validation prerequisite for any \polFF destined to perform MD simulations on large chemical and biochemical complexes.
\cite{AMOEBA03,Drudewater,TTM3-F,Liu2019, Liu2020,MBPOL,GEM}

The description of water with a \polFF is challenging \cite{paesanireview} as it requires a detailed understanding of its intermolecular interactions, and is also  a marker of its predictive ability. 
Several polarizable water models were recently reported (not all suitable for MD simulations), grounded either on Energy Decomposition Analysis (EDA) methods, such as the \mbucb \cite{Das2019} water model using the Absolutely Delocalized Molecular Orbital (\almo) \cite{Khaliullin2007}, or grounded on Symmetry Adapted Perturbation Theory (\sapt) \cite{Jeziorski1994,Parker2014} such as EFP (Effective Fragment Potential)  \cite{GORDON2007177,EFP}, the polarizable SAPT-5,\cite{SAPT5s} \mbpol \cite{Reddy2016}, Distributed Point Polarizable model (DPP) \cite{DPP}, Gaussian Electrostatic Model (\gem) \cite{GEM0,GEM,Gemstar,SAPTSehr}, Hydrogen-like Intermolecular Polarizable POtential (HIPPO\cite{HIPPO}), Derived Intermolecular Force-Field (DIFF) \cite{DIFF}, \amoebap (Atomic Multipole Optimized Energetics for Biomolecular Simulation +), \cite{Liu2019,Liu2020} and others.

The SIBFA \cite{NohadGresh2007} (Sum of Interactions Between Fragments \textit{Ab initio} Computed) polarizable force field is one of the very first ones to have resorted to Energy Decomposition Analysis (EDA) methods for its calibration. \cite{Gresh1982,Gresh1986,Gresh1995,NohadGresh2007} 
Such studies have shown the ability of SIBFA to reproduce the individual EDA contributions.
They were extended beyond Hartree--Fock to achieve the full separability of the potential \cite{JPPseparability2007} based on Constrained Space Orbital Variations (CSOV) DFT-based EDA reference computations.\cite{CSOV} 
However, even if it was introduced much earlier than the more popular  AMOEBA\cite{Ren2003,Laury2015} \polFF, the SIBFA interaction potential was also significantly more complex. Consequently, its applications to MD simulations were limited so far by the unavailability of analytical gradients for the entirety of its contributions, but also by the absence of an efficient periodic boundary conditions implementation.

In this work, such limitations have now been removed. Indeed, the SIBFA energy potential and its associated analytical gradients have been integrated into the massively parallel Tinker-HP molecular dynamics (MD) package \cite{Lagardere2018}, enabling the use of its native efficient Nlog(N) Smooth Particle Mesh Ewald implementation \cite{SPME,SPMELagardere} for multipoles and polarization.
This sets the ground for the first MD simulations with a full-fledged version of this potential since its inception. 

The article is organized as follows. First, we describe the functional form of the SIBFA potential and introduce a new and computationally tractable reformulation of the many-body charge transfer/delocalization energy. 
Then, as a proof of concept, we propose a prototypical SIBFA water model grounded on state-of-the-art SAPT(DFT)\cite{SAPTSehr}. Finally, we present the first complete set of SIBFA condensed-phase molecular dynamics simulations and discuss the computed water properties.

\section{Method}
\subsection{Overview of the SIBFA Functional Form}

The present SIBFA water model functional form embodies a refined many-body polarizable intermolecular potential designed to match its \abinitio counterpart obtained at the \saptdft level.\cite{misquitta_saptdft_2005,SAPTSehr} 
To allow the SIBFA model to be fully flexible, its functional form also includes an intramolecular potential and valence terms that are similar to the one used in the AMOEBA \polFF \cite{AMOEBA03}.
\subsubsection{Intramolecular Potential}
Following AMOEBA, the anharmonic bond-stretching and
anharmonic angle-bending are similar to the MM3 force field \cite{MM3}. Indeed, this confers to SIBFA anharmonicity using higher-order deviations from ideal bond lengths and angles. 
The coupling between stretching and bending modes is performed using an additional valence term, whereas an Urey-Bradley contribution \cite{urey1931vibrations} describes the water intramolecular geometry and vibrations. 
It is important to note that, in AMOEBA, the ideal bond length was chosen to be at the experimental value of $0.9572$ \AA, the ideal bond angle to $108.5^\circ$, and the Urey-Bradley ideal distance was set at $1.5326$ \AA. 
Additionally, it has been shown that setting up the ideal bond angle to a larger value than the gas phase angle of $104.52^\circ$ is useful to reproduce the experimental average angle in liquid water.\cite{AMOEBA03}
\subsubsection{Intermolecular Potential}
The SIBFA intermolecular potential is a sum of five contributions:

\[E_{inter}=E_{elec+pen}+E_{exch-rep}+E_{pol}+E_{ct/deloc}+E_{disp,exch-disp} \addtag \]
The contributions are: multipolar electrostatics (up to quadrupole) augmented with short-range penetration (with corrections up to charge-quadrupole), short-range exchange-repulsion using a $S^2$/$R$ formulation and expressed as a sum of bond-bond, bond-lone pair, and lone pair-lone pair components, many-body polarization, many-body charge-transfer/delocalization, and dispersion (up to C$_{10}$). 
The latter term also embodies an explicit exchange-dispersion term. 
The choice of the functional forms of the different contributions is motivated by the need to match their \abinitio SAPT(DFT) counterpart.

\subsection{Controlling the Accuracy of Short-Range Interactions}
The design of \abinitio molecular mechanics potentials grounded on quantum mechanics aims at reproducing accurately all types of interactions. In practice, it is of prime importance for these force fields to be able to capture short-range interactions that arise from regions of strong overlaps between molecular densities.
To do so, the natural strategy consists in building models based on explicit frozen electron densities such as GEM (Gaussian Electrostatic Model) \cite{GEM0,GGEM,statusGEM} or S/G-1 \cite{SG1}. These methods use resolution of identity techniques to derive high-quality model densities that can be used to generate explicit quantum chemistry analytic integrals. Such integrals enable the fast evaluation of density-based interactions with little approximation beyond the density discretization itself.
Such continuous electrostatics approaches are then able to include both long-range multipolar interactions and short-range overlap-dependent effects. 
These latter effects are needed to model electrostatic penetration, exchange-repulsion and accurate "quantum-like" electrostatic fields and potentials that will fuel the polarization/charge delocalization (i.e. induction) contribution.\cite{GEM0,SG1}

However, one challenge of this strategy is the need for accuracy for each one of the different contribution. Indeed each functional form does not have to rely on the same model electron density or to use it in the same way. 
In other words, errors do not impact each contribution with the same magnitude depending on the quality of the valence and/or core electronic density description. 

In practice, fitting density is a difficult problem and limits the accuracy and transferability of simple density models. 
Consequently, describing electrostatics (i.e. electron-electron plus electron-nucleus interactions) and overlap-dependent interactions using a single model density requires very high accuracy models. To ensure transferability and accuracy of electrostatics and overlap/repulsion contributions, it has been shown that multiple model densities could be important. \cite{GEM}

On the other hand, polarizable force fields such as SIBFA, or EFP (Effective Fragment Potential) \cite{GORDON2007177,EFP} mimic their short-range interactions using quantum-inspired damping functions. 
Indeed, their functional forms are grounded on simplified integrals obtained from Localized Molecular Orbital Theory. 
Managing error compensation and transferability is a key aspect of the design of such general approaches. 
SIBFA has its own strategy to address this fundamental problem. It uses a source of overlap specifically dedicated to electrostatics, whereas the others contributions use the overlap term originating from Exchange-Repulsion.
We detail in the next subsections the functional form of all intermolecular components, and an efficient reformulation of the many-body charge transfer/delocalization energy to ensure scalability. 

\subsection{Multipolar Electrostatics including Short-Range Penetration}
SIBFA models the electrostatic interactions through permanent multipoles up to quadrupoles. 
Dipoles and quadrupoles are stored in a local frame of the molecule they belong to, and are rotated in the global frame at each timestep. 
Multipoles are computed at the PBE0/aug-cc-pVTZ level (see parametrization discussion) using the Iterated Stockholder Atoms (ISA) approach.\cite{ISA}
The interactions between these permanent multipoles are screened at short-range to incorporate charge penetration effect. 
The damping function used is the one originally described in references \cite{Piquemal2003,JPPseparability2007}. 

\begin{equation}
   \begin{split}
    E_{elec}=E_{charge-charge*}+E_{charge-dipole*}+E_{charge-quadrupole*}+E_{dipole-dipole}+E_{dipole-quadrupole}+\\
    E_{quadrupole-quadrupole}
    \label{eq:sibfa-elec}
    \end{split}
\end{equation}
It is important to point out that the choice to come back to the original SIBFA formulation of the penetration component is motivated by its superior accuracy over a more recently introduced simplified variation. \cite{doi:10.1021/acs.jctc.5b00267,C6CP06017J}
We used here the full penetration correction scheme (i.e. noted with a * in \Eqrff{eq:sibfa-elec}) that includes charge-charge, charge-dipole and charge-quadrupole corrections.\cite{Piquemal2003,JPPseparability2007}. 
The coupling with the Smooth Particle Mesh Ewald (SPME) periodic boundary conditions approach is performed using the same correction strategy proposed in reference \cite{NarthJCC}. 

It is important to note that model parametrization is also designed to absorb the defects of multipolar electrostatics. 
Indeed, even if the ISA approach appears to be more robust than previous Distributed Multipole Analyses (DMA) \cite{DMA} approximations, a multipolar expansion is prone to errors when truncated at a given order (here at the quadrupole level). When adding the penetration correction and fitting to the reference SAPT(DFT) electrostatics, such errors are absorbed by the short-range damping function.
In practice, in order to avoid this spread to all overlap-dependent contributions the errors linked to the truncation of electrostatic multipoles, SIBFA does not intend to use electrostatics as its single source of overlap. Therefore, it relies on the specific overlap expressions present within the exchange-repulsion contribution to model all non-electrostatics overlap-dependent terms. It is possible since the present model focuses on modeling the valence only (by opposition to electrostatics that involves all electrons, see discussion in the next section).

\subsection{Exchange-Repulsion and anisotropy: the Importance of Explicit Lone-Pairs} 

The modeling of exchange-repulsion is of prime importance in SIBFA as its short-range overlap expressions are used in several other contributions.
Historically, the derivation of the functional form for exchange-repulsion terms is based on early theoretical results obtained by
Murrell et al. in references \cite{Murelloverlap1,Murrelloverlap2}
who proposed a simplified \abinitio perturbation scheme representing the exchange-repulsion energy based on an overlap expansion (S) of localized Molecular Orbitals (LMOs).  
Based on additional results by Salem and Longuet-Higgins \cite{Salem} highlighting the importance of the LMOs representation to model repulsion-like interaction and on expressions proposed in reference \cite{Gresh1986}, the exchange-repulsion can be expressed as a sum of valence-only terms, namely: bond-bond, bond-lone pair and lone pair-lone pair interactions such as:

\[E_{rep}=E_{rep, bond-bond}+E_{rep,lp-lp}+E_{rep,bond-lp} \addtag\]
The key feature is that the bond-bond, bond-lone pair and lone pair-lone pair overlaps are developed as a sum of atom-atom overlaps. This is based on the idea that a bond orbital can be decomposed into two valence s-type atom-centred orbitals with appropriate angular factors on the second orbital to take hybridization into account.\cite{Gresh1986,Gresh1995}

The exchange-repulsion formalism \cite{Gresh1986,Gresh1995} evolved over the years and the expressions used here are based on its latest iteration evolution that was reported in  \citet{robinexchange}. 
It uses explicit lone pair extra-centers. 
Their position was extracted from a DFT Boys Localization procedure \cite{Boys} at the PBE0/aug-cc-pVTZ, and automatically refined to minimize the SIBFA contribution errors with respect to the reference SAPT(DFT) data (the present positions are given in SI, see section 5). 
The flexibility of the approach also allows to design groups of smeared lone-pairs to overcome difficult systems where multipoles alone could not capture the full \abinitio reference anisotropy.\cite{khourysmeared} 
In the rest of the manuscript, all described overlap contributions and use of lone-pairs will be based on the present formalism. 
Full technical details about the formulation of the exchange-repulsion component can be found in the technical appendix.

\subsection{Many-body Polarization}

The polarization energy used here is similar to the AMOEBA water model: we use isotropic scalar polarizabilites and Thole damping \cite{THOLE1981341,AMOEBA03,scalablelip,SPMELagardere} at short-range. 
Computationally, the induced dipoles can be converged using the same machinery that is commonly used with AMOEBA within Tinker-HP.\cite{lagardere2015scalable,Aviat2017,Aviat2017a,pushing} 
Details can be found in the technical appendix.

\subsection{Many-Body Charge Transfer/Charge Delocalization}

As for the exchange-repulsion, the many-body Charge Transfer/Charge Delocalization energy is derived from an early simplified perturbation theory due to Murrel, Randić and Williams \cite{MurellTC}. It was then initially adapted by Gresh \etal \cite{Gresh1982,Gresh1986} to become applicable to force fields. 
In practice, the definition of this contribution is not straightforward and has led to many discussions in the literature. Charge Transfer/Charge Delocalization is mainly associated with the sharing or tunneling of the electrons of the interacting monomers onto the electron-deficient sites of the partners, resulting then in a lowering of the energy of the complex.
Thus, in line with our recent SAPT(DFT) study \cite{SAPTSehr} and following Misquitta, \cite{Misquitta13} we term this contribution Charge Transfer/Charge Delocalization. 
In the present Tinker-HP SIBFA implementation, two options are available. 
The first corresponds to the original equations that scale as N$^3$. 
We propose here a novel formulation of this contribution, lowering its complexity to ensure its applicability to realistic systems (see section below).

\subsubsection{Reducing the Complexity of the Many-Body Charge Transfer/Delocalization Energy}

The present advantage of the original formulation of the Charge Transfer/Charge Delocalization energy is its capability to capture many-body interactions. 
However, despite its elegance there is a computational price to pay for such a contribution as the formula scales as N$^3$ to get analytical gradients, limiting the possibility to perform molecular dynamics. 
We present here a novel resolution of the many-body charge-transfer/delocalization term that we formulated in order to diminish its computational impact.

The charge-transfer between an electron donor molecule A and an electron acceptor molecule B can be expressed as:
\[E_{ct}=-2\sum_{\alpha}\sum_{\beta^*}\frac{I_{\alpha\beta^*}^2}{\Delta E_{\alpha\beta^*}} \addtag \]
where $\alpha$ are the occupied molecular orbitals of A and $\beta^*$ are the unoccupied molecular orbitals of B. 
$\Delta E_{\alpha\beta^*}$ is the energy associated to the electron transfer between $\beta^*$ and $\alpha$ and
\[I_{\alpha\beta^*}=\int \rho_{\alpha\beta^*}(r)V(r)dv \addtag \]
and where $\rho_{\alpha\beta^*}$ is the overlap transition density:
\[\rho_{\alpha\beta^*}=-(\alpha\beta^*-\alpha^2S_{\alpha\beta^*}) \addtag \]
where $S_{\alpha\beta^*}$ is the overlap integral between $\alpha$ and $\beta^*$. The derivation of an operational formula of \ect starting from these equations is given in Refs.\cite{Gresh1982,Gresh1986,Gresh1995} 

In SIBFA, the electron donors are the lone-pairs of the molecules and the electron acceptors are all H-X bonds, where X denotes a heavy atom. 
For each lone pair $i_{lp}$, let us denote by $c(i_{lp})$ the index of its carrier, by $V_{i_{lp}}$ the electrostatic potential on $i_{lp}$ due to the complete system except the molecule it belongs to. 
For a given acceptor $k$, let us denote by $a_1(k)$ and $a_2(k)$ the indexes of both the atoms constituting it, $a_1(k)$ being the heavy one and $a_2(k)$ being the hydrogen one. 
We call $V_{a_1(k)}$ and $V_{a_2(k)}$ the electrostatic potential on $a_1(k)$, and $a_2(k)$ due to the complete system (except the molecule they belong to), and when considering an donor-acceptor pair $k-ilp$ by $V_{a_1(k)}*_{ilp}$ and $V_{a_2(k)}*_{ilp}$ the same electrostatic potential minus the contribution of the molecule $i_{lp}$ belongs to.
We approximate the intramolecular electrostatic potential on $a_1(k)$ and $a_2(k)$, $V_{int}(a_1(k))$ and $V_{int}(a_2(k))$ by using the expression of equation 24 and 25 of reference\cite{Gresh1986}. 
We denote by $n(i_{lp})$ the occupation number of the lone pair $i_{lp}$, we assign van der Waals radii to the lone pair carrier $r_{vdw}i$, and the two atoms constituting the acceptor: $r_{vdw}a_1(k)$, $r_{vdw}a_1(k)$. We then define the two reduced distances as:
\[\rho_{1}=\frac{r_{c(i_{lp})a_1(k)}}{2\sqrt{r_{vdw}ir_{vdw}a_1(k)}} \addtag\]
\[\rho_{2}=\frac{r_{c(i_{lp})a_2(k)}}{2\sqrt{r_{vdw}ir_{vdw}a_2(k)}} \addtag\]
where $r_{c(i_{lp})a_1(k)}$ is the distance between the lone pair carrier and the first atom of the acceptor, and $r_{c(i_{lp})a_2(k)}$ the distance between the lone pair carrier and the second atom of the acceptor.

For each donor-acceptor pair, we define the following exponential terms:
\[e_1=e^{-\eta \rho_1}(V_{int}(a_1(k))+V_{a_1(k)}*_{ilp}-V_{i_{lp}})(C_st_{i_{lp}a_1(k)}+C_pm_{i_{lp}a_1(k)}cos(\alpha)) \addtag \]
where $\eta$ is a parameter, $C_s$ and $C_p$ hybridization coefficient, $t_{i_{lp}a_1(k)}$ and $m_{i_{lp}a_1(k)}$ tabulated coefficient used to approximate integrals described in reference 24 and 25, and $\alpha$ the angle formed by the two vectors: $r_{c(i_{lp})i_{lp}}$ and $r_{c(i_{lp})a_1(k)}$,
\[e_2=e^{-\eta \rho_2}(V_{int}(a_2(k))+V_{a_2(k)}*_{ilp}-V_{i_{lp}})(C_st_{i_{lp}a_2(k)}+C_pm_{i_{lp}a_2(k)}cos(\beta)) \addtag \]
with the same tabulated coefficients as for the $e_1$ term, and with $\beta$ the angle formed by the two vectors: $r_{c(i_{lp})i_{lp}}$ and $r_{c(i_{lp})a_2(k)}$.

The final charge transfer energy used in SIBFA for water then reads:\[E_{ct}=\sum_{i_{lp}}\sum_{k}-n(i_{lp})\frac{(e_1-e_2)^2}{\Delta E} \addtag \]
with 
\[\Delta E=ae(i)+V_{i_{lp}}-(ah((a_1(k))+V_{a_1(k)})\addtag \]
where $ah(a_1(k))$ is the electronic affinity of $a_1(k)$, and $ae(i)$ the ionization potential of the lone pair.
In practice, the total electrostatic potential on the acceptors is precomputed in a first stage as well as the total electrostatic potential on the lone-pairs.
Then, each contribution is computed through a double loop among the donors (lone-pairs) and the acceptors (X-H bonds). 
In particular, $V_{a_1(k)}*_{ilp}$ and $V_{a_2(k)}*_{ilp}$ are obtained by removing the contribution from the local molecule of $ilp$ to the precomputed electrostatic potential, which has a bounded computational cost (with respect to the number of atoms of the system). 
Finally, to obtain, $E_{ct}$, $O(N^2)$ operations are needed to precompute the potential, and $O(N^2)$ operations to compute the interactions through the double loop. 
To compute analytical gradients of $E_{ct}$, one needs first to precompute the derivatives of the electrostatic potentials (and also take into account the derivatives associated with the rotation matrices to get the permanent multipoles in the global frame), which requires also $O(N^2)$ operations. 
Then, for each of the donor-acceptor pair, one needs a bounded number of operations to get the derivatives of $e_1$ and $e_2$, and additional operations to remove the contributions due to the molecule to which $ilp$ belongs. 
But one needs to explicitly sum these contributions to the gradients with respect to all the atoms (that contribute to the potentials) adding a (small) $O(N)$ number of operations per pair for a final $O(N^3)$ computational cost. 

Nevertheless the actual computational cost lies more in the initial computation on the electrostatic potentials and of their derivatives so one would like to simplify these expressions. The most straightforward way to simplify these terms for the acceptors in the $e_1$ and $e_2$ expressions is to reduce them to $V_{int}(a_1(k))$ and $V_{int}(a_2(k))$, such a simplification does not exist for the potential on the lone-pairs $V_{i_{lp}}$.
We then chose to modify further the formulation of $E_{ct}$ so that $V_{a_1(k)}$ is removed from $\Delta E$ and so that the potentials $V_{a_1(k)}*_{ilp}$ and $V_{a_2(k)}*_{ilp}$ are not considered, yielding a computationally much cheaper expression. 
By using cutoffs and neighbor lists, the scaling decreases to $O(N^2)$ making then the formula compatible with MD simulations. 
The reduced expression we used then reads:
\[E_{ct}=\sum_{i_{lp}}\sum_{k}-n(i_{lp})\frac{(e_1-e_2)^2}{\Delta E} \addtag \]
with 
\[e_1=e^{-\eta \rho_1}(V_{int}(a_1(k))-V_{i_{lp}})(C_st_{i_{lp}a_1(k)}+C_pm_{i_{lp}a_1(k)}cos(\alpha)) \addtag \]
\[e_2=e^{-\eta \rho_2}(V_{int}(a_2(k))-V_{i_{lp}})(C_st_{i_{lp}a_2(k)}+C_pm_{i_{lp}a_2(k)}cos(\beta)) \addtag \]
\[\Delta E=ae(i)+V_{i_{lp}}-ah(a_1(k))\addtag \]

\subsection{Dispersion and Exchange-Dispersion}
The dispersion model used in SIBFA includes both a dispersion encompassing C$_{6}$, C$_{8}$ and C$_{10}$ coefficients, and an exchange-dispersion term. 
Since the overlap-dependent exchange-repulsion is present, the global expression involves atoms and lone-pairs.
More precisely, it contains an atom-atom term, a lone-pair-lone-pair term and a cross atom-lone pair term. 
A detailed description of the formalism of the contribution can be found in the technical appendix.

\section{Strategy for Gradients Evaluation and Associated Computational Complexity}
One of the key achievements of this work is to provide all the necessary analytical gradients to perform molecular dynamics. 
The detailed strategy to compute the analytical gradients of $E_{ct}$ has been described above, and the strategy to efficiently compute the polarization energy and associated forces has been described elsewhere\cite{scalablelip,lagardere2015scalable}. 
The detailed expression of both the exchange-repulsion and dispersion contributions used in our model is presented in the technical appendix. 
In short, for both the dispersion and the exchange repulsion energy, a careful derivation of all the terms involved allows for a naive evaluation of the gradients with a quadratic complexity. 
In the case of exchange-repulsion, this can be made linear with respect to the number of atoms through the use of cutoffs and neighbor lists. 
This is not the case for dispersion due to its intrinsically long-range nature.

\section{Calibration of the Intermolecular Potential: from SAPT(DFT) to \ccsdt}
One key aspect of the SIBFA design is to rely only on \abinitio data extracted from a Born-Oppenheimer reference surface.
The choice of the reference computations is therefore crucial. 
Indeed, since the SIBFA potential relies on the derivation of ISA distributed multipoles, the potential is linked to their level of extraction. 
In the same connection, SIBFA relies also on SAPT(DFT) \cite{SAPTSehr} computations to compute the different reference contributions. 
Therefore, the PBE0/aug-cc-pVTZ level is our reference for the derivation of multipoles and SAPT(DFT) computations.

However, SAPT(DFT) energies tend to underestimate \ccsdt ones (see Table 1). Thus, the final parametrization should also aim to correct this by being as close as possible to the \ccsdt reference.
The present SAPT(DFT) version has been chosen as it is able to provide accurate exchange-repulsion contribution energies (it uses the full expression and not the usual \Ssq approximation) that have been shown to be also important to obtain meaningful values of the polarization and charge transfer contributions (see reference \cite{SAPTSehr} for details), an important feature to achieve the separability of the SIBFA potential. 
Therefore, our parametrization is a two-step procedure.

First, we perform a simple but automated optimization of the parameters of each contribution on a deliberately limited set of $39$ water dimers (see Figure S1). 
Such computational experiment is designed to test the capability of the SIBFA equations to transfer towards condensed phase without the support of an extensive data-driven strategy. 
For each contribution, using the BFGS procedure \cite{hsl}, all relevant parameters were optimized simultaneously by minimizing the square root of the summed SIBFA vs. \saptdft differences.  
Second, owing to the limited number of dimers in the training set, and to the present absence of the nuclear quantum effects at play in the present MD simulations, it was necessary to do a limited recalibration of four parameters in order to minimize further the difference with \ccsdt results while improving the transition to the liquid phase. In that regard, beyond the CCSDT energies, we included two simple experimental properties at $298$K: the density and the position of the first peek of the O-O radial distribution function (RDF). While little affecting the individual contributions calibrated on SAPT(DFT), this additional steps overall improved the condensed phase properties (see final values on Tables 1 and S1). 

The four modified parameters are: the vdw radius of the O atoms used for \erep, decreased from $1.5077$ to $1.4755$; the corresponding vdw for \edisp, passing from $1.2644$ to $1.2594$; the corresponding vdw on the fictitious atom used for \edisp, located on the tip of each sp$^3$ lone-pair, reduced from $0.9542$ to $0.9042$; and the Thole damping used for \epol, decreased from $0.39$ to $0.28$. 
Condensed phase properties were derived at the outcome of MD over a range of temperatures (from $261$K to $369$K) in order to validate the \sibfa water model against experimental data. 
This final version is denoted \sibfa and no further optimization was performed.
All final parameters, multipoles and polarizabilities can be found in SI (see section S5).

\section{Results and Discussion}

\subsection{Achieving the Potential Separability and Agreement with SAPT(DFT)}

\begin{figure}[H]

\begin{center}

\includegraphics[viewport=70 90 1000 1190, clip, scale=0.52]{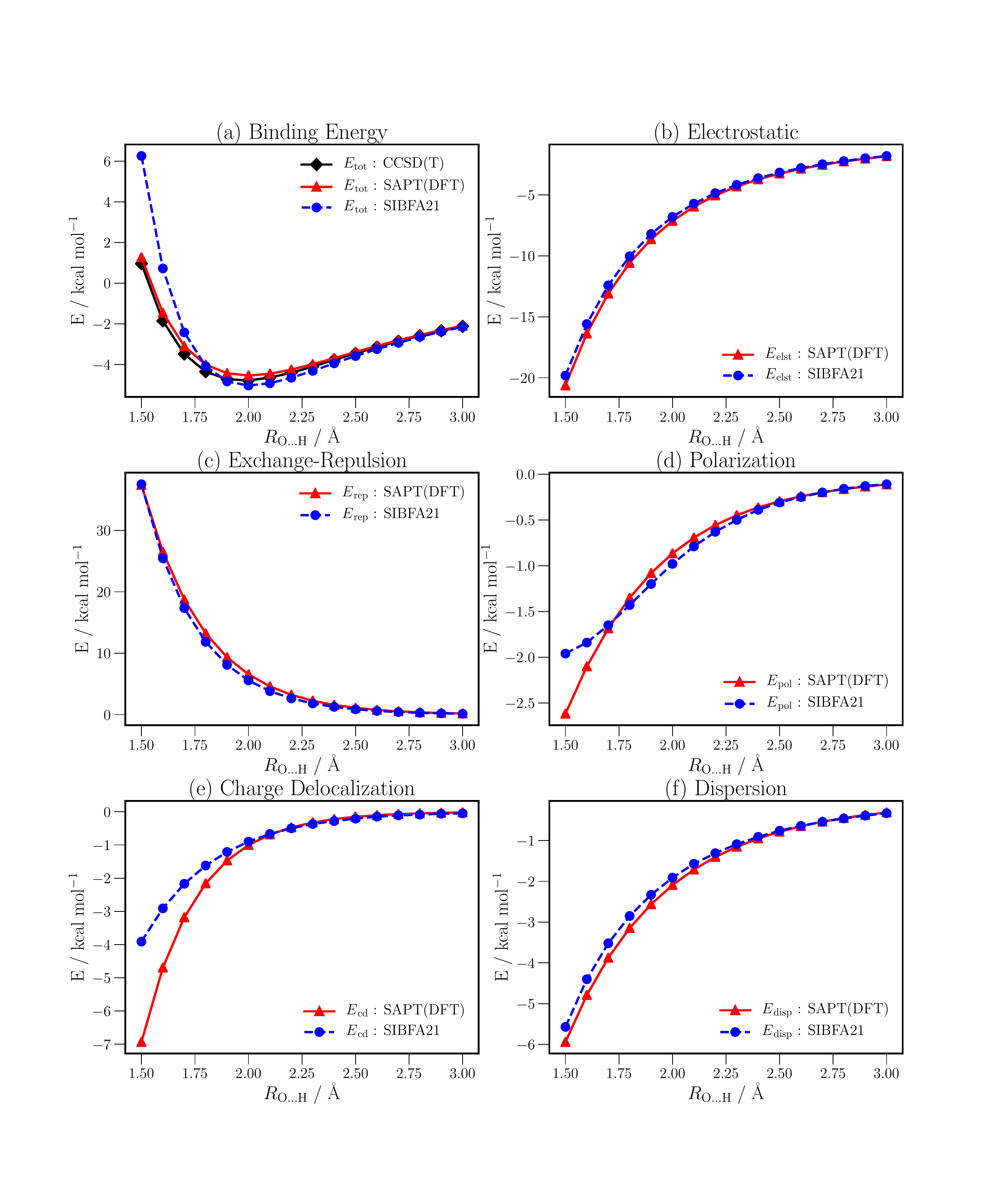}

\caption{ 
Comparison of \sibfa and \saptdft energy components for the linear water dimer. 
The PBE0 functional and the aug-cc-pVTZ basis were used in \saptdft calculations.
}
\label{fig:SIBFA_SAPTDFT_WATER_LINEAR}
\end{center}
\end{figure}

\figrff{fig:SIBFA_SAPTDFT_WATER_LINEAR} represents for the linear water dimer the radial evolution of \sibfa binding energy and of its five contributions, for variations of the O\dots H distance from $1.5$ \AA \xspace to $3.0$ \AA.  
The Root Mean Square Error (RMSE) for the electrostatic, polarization, and dispersion energies are less than $0.4$ \kcalmol, and less than $1$ \kcalmol for the exchange-repulsion and charge-delocalization energies. 
Close RMSE values and radial behaviors obtained with \sibfa indicates that the gas phase interaction energies and their separate contributions have not been altered by the limited adjustment of the parameters based \ccsdt-reference/ condensed phase properties. 
Notably, the effects at very short-range are correctly captured regarding the proper separability of the second-order energy into polarization and charge-delocalization (see also Table S1).

We next compare the binding energies of the Smith dimers obtained with \sibfa (\tabrff{tab-sibfa-vs-saptdft-smith-dimers} and \tabrff{tab-ebind}). 
Such energies are the sum of the intermolecular and intramolecular (deformation) energies. 
Here, we also set the HOH angle at $108.5^{\circ}$ which is close to the condensed phase experimental value, whereas in gas phase the HOH angle is $104.52^{\circ}$. 
As mentioned above, it has been shown that the choice of a larger angle value is required to accurately represent the geometry fluctuation of water in condensed phase. \cite{Ren2003} 
The present model does not rely on charge flux as for AMOEBA+ \cite{AMOEBA+2} in order to test the intrinsic capabilities of the native SIBFA components. 
Moreover, the \sibfa water model is flexible leading to a good agreement of binding energy for the Smith dimers. 
The RMSE from \sibfa compared to \ccsdt/CBS is around $0.6$ \kcalmol comparable to \amoeba and \mbucb (see Table S4). 
But the result is in closer agreement with the \ccsdt/aug-cc-pVTZ level that exhibits itself a 0.26\kcalmol with the CBS level (see Table S4). The SIBFA results were expected since both reference SAPT(DFT) data and ISA multipoles are not compute at a CBS level but obtained with the aug-cc-pVTZ incomplete basis set. 
Overall, \sibfa appears capable of correctly describing attractive and repulsive interactions at short- and long-range in the gas phase.

At this point, one could say that these results probably illustrates some  limits of a classical potential. 
Indeed, it has been shown that the importance of triple excitations of the various Smith dimers, range already from $0.06$ to $0.2$ \kcalmol when using the aug-cc-pVTZ basis set, and should be even larger at the complete basis set limit (CBS).\cite{ReinhPiqwater} 
Such high-resolution correlation/overlap dependent interactions being missed by post--Hartree Fock methods such as MP2 \cite{ReinhPiqwater} seem particularly difficult to capture for any classical potential and should be carefully addressed to avoid unrealistic overfitting. 
SIBFA reaches here a quality close to the electron density based GEM-0 interaction potential \cite{GEM0} which is satisfactory.
\subsection{Transferability of the SIBFA21 potential} 
We evaluate the accuracy of the present potential upon passing from water dimer to water clusters with $n=3$ up to $n=20$ molecules. 
We compare the numerical values and trends of the \sibfa binding energies (\ebind) to the available high-level \abinitio reference data obtained with extended basis sets (\tabrff{tab-ebind}). 
For the hexamers and endecamers, the RMSE is about $1.5$ \kcalmol, the relative error being always less than $3$ percent. Overall, for hexamers, the agreement is satisfactory with only a minor inversion in two hexamers, 5, and 6, that have close binding energies.
For the 16-mer and 20-mer, the RMSE is slightly larger, amounting to $2$ and $5$ \kcalmol respectively. 
Despite the much larger magnitudes of \ebind, the relative errors remain $<2$ percent, with one exception, that of the fused cube 20-mer, for which it amounts to $4$ percent.
For all water clusters ($n=3$ to $n=20$), the RMSE computed with \sibfa amounts to $2.5$ \kcalmol, virtually equal to the \mbucb one, but smaller than the $5.5$ and $6.3$ \kcalmol found with \amoeba and \amoebap, respectively (see Table S2 and Figure S2). 

\subsection{Condensed Phase Molecular Dynamics Simulations} 

\begin{figure}[H]

\begin{center}
  \includegraphics[viewport=10 40 600 1200, clip, scale=0.52]{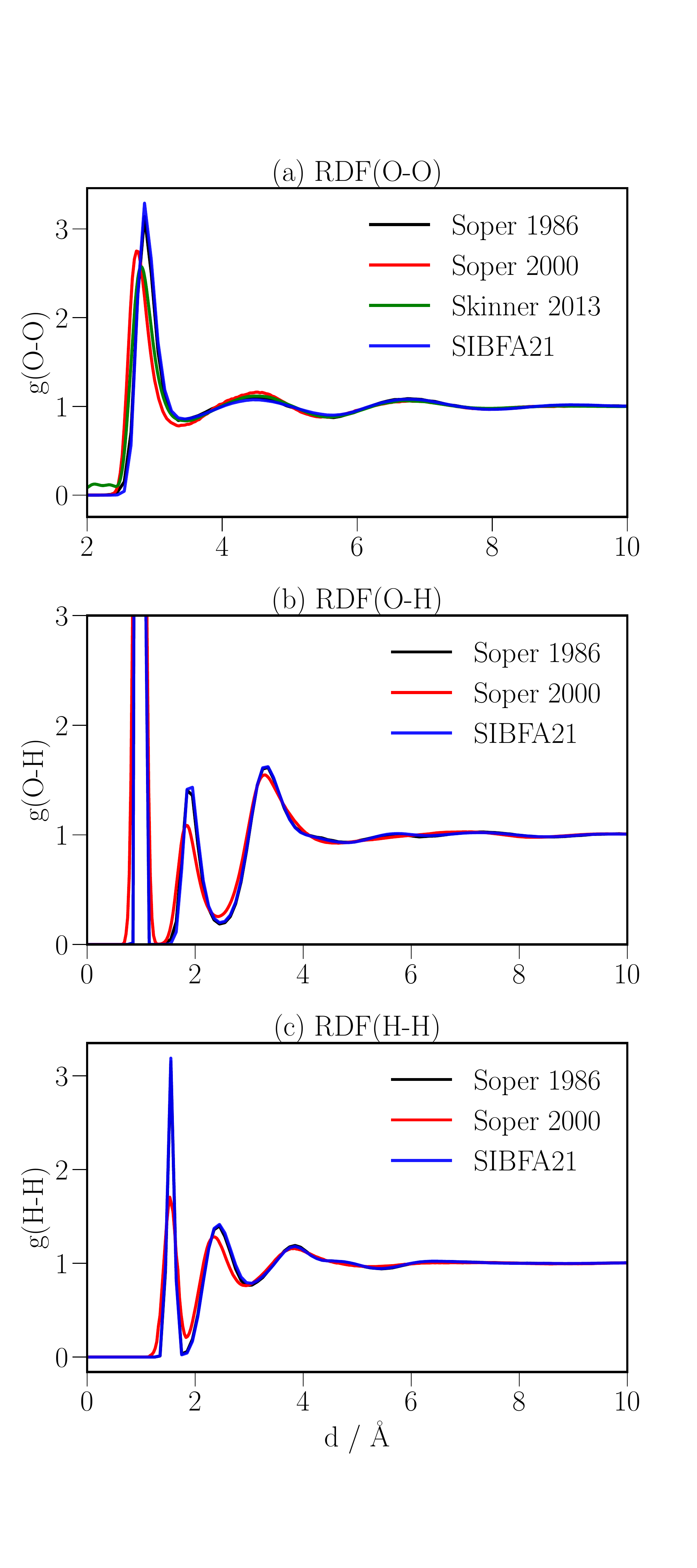}

\caption{ 
Comparison of the radial distribution functions (O-O), (O-H) and (H-H) computed with \sibfa to the experiment at 298K.
}
\label{fig:SIBFA_WATER_RDF}
\end{center}
\end{figure}

The radial distribution function (RDF) of water at $298$K is represented in \figrff{fig:SIBFA_WATER_RDF}.
The RDF(O-O) is in very good agreement with the experimental values reported by Soper \etal \cite{Soper1986,Soper2000} and Skinner \etal \cite{Skinner2013}, as well as for the RDF pairs O-H and H-H. 
Thus, the location and the heights of the first O-O, O-H and H-H peaks can closely match the experimental ones. This is also the case for the secondary peaks and for the entirety of the three curves. 
This is a convincing indication that the first and second solvation shells of water in the liquid phase can be correctly determined with the present \sibfa calibration.

\begin{figure}[H]

\begin{center}

  \includegraphics[viewport=70 160 1300 1550, clip, scale=0.45]{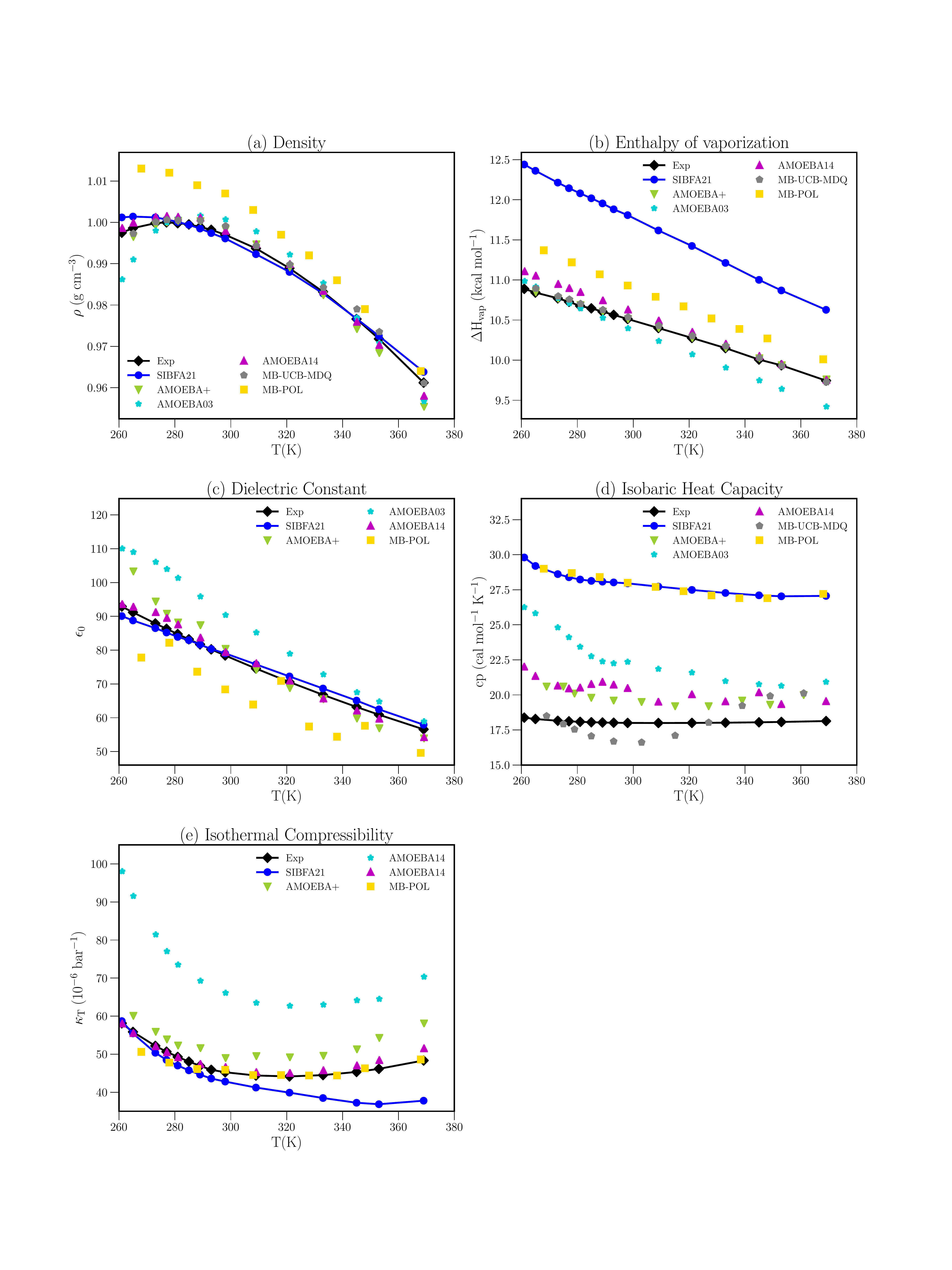}

\caption{ 
Comparison of condensed phase properties (from 261K to 369K) computed with \sibfa, \amoebap, \amoeba, \mbpol and \mbucb to the experiment.
}
\label{fig:PROPERTIES_SIBFA}
\end{center}
\end{figure}

We report also in \figrff{fig:PROPERTIES_SIBFA} the main condensed phase properties of water computed at different temperature from $261$K to $369$K. 
We observe that while the density (\dens), the dielectric constant (\dielec), and the isothermal compressiblity (\isocomp) are closely reproduced, the enthalpy of vaporisation (\vap) and the isobaric heat capacity (\heat) are overestimated. Whereas, the self-diffusion constant (\diff) is underestimated. 
At $298$K, the present $1.3$ \kcalmol overestimation of \vap by \sibfa (11.8 \kcalmol) is not surprising and appears in the range of the values obtained by previous polarizable MD simulations studies lacking Nuclear Quantum Effects (NQEs).\cite{Fanourgakis2006,Paesani2007} 

The high overestimation of \heat (28 \ecp) is a consequence of the overestimation of \vap. Indeed, a larger magnitude of the latter is reflected by a larger heat capacity. Overalll, it has been shown that \heat and \vap are more subject to overestimation if the NQE are not included. 
Moreover, Levitt \etal \cite{Levitt1997} have suggested to subtract $6$ \ecp to correct the MD value. 
Such a correction enables a closer agreement of the \sibfa \heat with the experimental value ($18$ \ecp). 
The underestimated self-diffusion constant ($1.47$ \ediff) on the other hand is a reflection of the slower rearrangements of the water-water H-bonds and larger frictional forces.
Finally, the self-diffusion constant has been shown to be dependent of the size of the box. 
Using a correction proposed in Ref \cite{Yeh2004}, the \sibfa \diff increases now by $0.4$ \ediff, getting then closer to experiment ($2.29$ \ediff).

\section{Conclusion and Perspectives} 
We have reported the first MD simulations of liquid water using the polarizable, multipolar SIBFA many-body potential. 
This was enabled by the simplification of the many-body charge transfer contribution, and by the derivation and implementation of its analytical gradients in the massively parallel \tinkerhp code. 
An essential asset is the prior automated two-step calibration procedure of the parameters of its five contributions so that each matches its counterpart from \saptdft but also, overall minimizes the error with a reference \ccsdt level while yielding to improved condensed phase properties. 
We have deliberately used a training set limited to $39$ water dimers as well as a simple parametrization procedure.
As opposed to the strategy used by most other \polFFs, we did not resort to extensive recalibration of all parameters to reproduce both cluster \abinitio \ebind results, and experimental observables from the condensed phase.
Therefore, we have demonstrated here the capability of the quantum-inspired SIBFA model to predict satisfactory gas-phase and condensed phase properties thanks to the physical motivation of each of its components. 

For example, the binding energy computed with SIBFA is in close agreement with \abinitio reference, with relative errors generally $<2$ percent for water dimers and clusters.

The next step is to refine further the approach using a full-fledged data-driven strategy to obtain a production water model. 
This is currently in progress.
This work opens a new avenue for condensed phase simulations of complex systems as both the non-additivity and the non-isotropy of the SIBFA potential were amply validated in numerous previous studies from our Laboratories. \cite{gresh1997,guo2000,antony2005,Roux2007,piquemal2007key,NohadGresh2007,bimetallic,goldwaser2014conformational,addressing,Devillers2020,darazi2020}
This should ensure for a safe transferability, from the training set data to proteins and nucleic acids.
At this stage, it is important to note that the simple AMOEBA intramolecular potential used in this study is not optimal since its ideal bond angle was set to $108.5^\circ$. 
However, based on the work performed by some of us, it would be straightforward to upgrade it with a geometry dependent charge flux (GDCF) model to follow the choices of AMOEBA+ \cite{Liu2020}. 
Still, we can only stress now the need to explicitly include Nuclear Quantum Effects (NQEs) \cite{paesanireview}. 
In that connection, we proposed here an implementation of the "canonical" SIBFA potential and the potential will continue to evolve  \cite{poier_lagardere_piquemal_2021}. The possible extensions of the functional form will therefore require an in-depth study of the role of NQEs to avoid double counting and to control many-body effects.
Indeed, as we mentioned, we have intended here to only ground the SIBFA potential on Born-Oppenheimer surface reference \abinitio data: explicit NQEs are missing in our present simulations. 
 
Forthcoming studies integrating NQEs in the SIBFA simulations will be necessary to evaluate their impact on various observables at various temperatures. Coupling of SIBFA to NQE methods (i.e. fast approaches such as Quantum Thermal Bath (QTB) \cite{mauger2021}, path integrals...) and its portage to GPUS\cite{GPU} to benefit from further HPC acceleration are therefore mandatory and are in progress too.  

\section{Computational Details}
All MD simulations were performed using the CPU implementation of Tinker-HP \cite{Lagardere2018} in double precision. The Tinker-HP SIBFA implementation is freely accessible to academic users through the general Tinker-HP GitHub repository \cite{THPGithub}.
Periodic boundary conditions were applied within the framework of Smooth Particle Mesh Ewald summation, with a grid of dimensions 120x120x120 using a cubic box with side lengths of $18.643$ \AA. 
The Ewald cutoff was set to $7$ \AA, the exchange-repulsion cutoff was $7$ \AA, \xspace and dispersion cutoff was $20$ \AA.
Molecular dynamics simulations were performed using the Beeman integrator ($1$ fs timestep), a preconditioned conjugate gradient polarization solver (with a 10$^{-5}$ convergence threshold) to solve polarization at each time step\cite{lagardere2015scalable}.
MD was performed in a water box of $216$ molecules in the NPT ensemble during $15$ns of production. 
Additional quantum chemistry computations were performed with GAUSSIAN09 \cite{frisch2009g09}, as all SAPT(DFT) calculations used CamCASP \cite{misquitta_camcasp}.
All properties were computed following the procedures described in SI (see section S4).

\section{Technical Appendix}
\subsection{Description of the Exchange-Repulsion Formalism}
As already described in \citet{robinexchange}, the exchange-repulsion model of SIBFA is overlap-based and involves interactions between bonds and bonds, lone-pairs and lone-pairs and cross interactions between bonds and lone-pairs.
\[E_{rep}=E_{rep, bond-bond}+E_{rep,lp-lp}+E_{rep,bond-lp}\]
which can be decomposed in pairwise interactions such that for example:
\[E_{rep,bond-bond}=\sum_{i,j=1,i < j}^{nbonds}E_{i,j,rep, bond-bond}\]
For each of these pairs, one has to build the overlap between the orbitals of the involved bonds which can be expressed as a combination of overlap between $s$ and $p$ orbitals as written in \citet{Gresh1986}. Let us denote by $a$ and $b$ the atoms forming the $i$-th bond and by $c$ and $d$ the ones forming the $j$-th bond. Each atom of each bond bearing a $s$ and a $p$ orbital, the overlap between the two bonds can be decomposed in 4 terms involving atomic pairs, each of them in term being made of four terms involving $s$ and $p$ orbitals, such that the overlaps are:
\[O_{ac}=O_{s_a s_c}+O_{s_a p_c}+O_{s_c p_a}+O_{p_a p_c}\]
\[O_{ad}=O_{s_a s_d}+O_{s_a p_d}+O_{s_d p_a}+O_{p_a p_d}\]
\[O_{bc}=O_{s_b s_c}+O_{s_b p_c}+O_{s_c p_b}+O_{p_b p_c}\]
\[O_{bd}=O_{s_b s_d}+O_{s_b p_d}+O_{s_d p_b}+O_{p_b p_d}\]
The term involving $s$ orbitals consist in the product of hybridation coefficents such that for example:
\[O_{s_a s_c}=C_{s_a} C_{s_c}\]
The terms involving $s$ and $p$ orbitals read:
\[O_{s_a p_c}=C_{s_a}C_{p_c}f_{ac}cos(\vv{AB},\vv{AC})\]
\[O_{s_c p_a}=C_{s_c}C_{p_a}f_{ca}cos(\vv{CD},\vv{CA})\]
with $f_{ab}$ and $f_{ca}$ tabulated values to approximate the corresponding integral, and the ones involving both $p$ orbitals read, for example:
\[O_{p_a p_c}=C_{p_a}C_{p_c}2cos(\vv{AB},\vv{AC})(\vv{CD},\vv{CA})\]

These overlap terms are to be multiplied by exponential prefactors:
\[S_{ac}=M_{ac}e^{-\alpha \rho_{ac}}\]
\[S_{ac_2}=M_{ac}e^{-\alpha_2 \rho_{ac}}\]
with $\alpha$ and $\alpha_2$ two constants and 
\[M_{ac}^2=g_a g_c (1-\frac{Q_a}{N_a^{val}})(1-\frac{Q_c}{N_c^{val}})\]
with $g_a$ and $g_c$ parameters, $Q_a$ and $Q_c$ the partial charges of atoms $a$ and $c$, $N_a^{val}$ and $N_b^{val}$ the number of valence electrons of atoms $a$ and $b$. The reduced distance $\rho_{ac}$ is defined as:
\[\rho_{ac}=\frac{r_{ac}}{4\sqrt{r_{vdw_a}r_{vdw_c}}}\]
with $r_{vdw_a}$ and $r_{vdw_b}$ distances specific to atoms $a$ and $b$. Similarly, we can define $S_{ad}$, $S_{ad_2}$, $S_{bc}$, $S_{bc_2}$, $S_{bd}$, $S_{bd_2}$.
Finally let us call:
\[O_{1_{i,j}}=O_{ac}S_{ac}+O_{ad}S_{ad}+O_{bc}S_{bc}+O_{bd}S_{bd}\]
\[O_{2_{i,j}}=O_{ac}S_{ac_2}+O_{ad}S_{ad_2}+O_{bc}S_{bc_2}+O_{bd}S_{bd_2}\]
Then if we define call $\vv{r_{mid_{ab}}}$ the position vector of the midpoint of $a$ and $b$ and $\vv{r_{mid_{cd}}}$ the position vector of the midpoint of $c$ and $d$, if we write
\[\vv{r_{mid}}=\vv{r_{mid_{ab}}}-\vv{r_{mid_{cd}}}\]
we have
\[E_{i,j,rep, bond-bond}=Occ_{ab}Occ_{bd}(C_1\frac{O_{1_{i,j}}^2}{r_{mid}}+C_2\frac{O_{2_{i,j}}^2}{r_{mid}^2})\]
with $C_1$ and $C_2$ constants and $Occ_{ab}$ and $Occ_{bd}$ the bond occupation number of bond $i$ and $j$

Similarly, we have:
\[E_{rep,lp-lp}=\sum_{i,j=1,i < j}^{nlp}E_{i,j,rep-lp-lp}\]
we can define an overlap term as:
\[O_{ij}=O_{s_i s_j}+O_{s_i p_j}+O_{s_j p_i}+O_{p_i p_j}\]
Let us denote by $a$ the atom carrying the lone pair $i$ and $b$ the atom carrying the lone pair $j$. As for the bond-bond part, the term involving s orbitals consists in the product of hybridation coefficients (of the atom carrying the lone-pairs), such that:
\[O_{s_i s_j}=C_{s_a}C_{s_b}\]
The terms involving $s$ and $p$ orbitals read:
\[O_{s_i p_j}=C_{s_i}C_{p_j}f_{ab}cos(\vv{BJ},\vv{BI})\]
\[O_{s_j p_i}=C_{s_j}C_{p_i}f_{ab}cos(\vv{AI},\vv{AJ})\]
The term involving both $p$ orbitals reads:
\[O_{p_i p_j}=C_{p_a}C_{p_b}2cos(\vv{AI},\vv{AJ})(\vv{BJ},\vv{BI})\]
We also have exponential prefactors:
\[S_{ij}=M_{ij}e^{-\alpha\rho_{ij}}\]
\[S_{ij_2}=M_{ij}e^{-\alpha_2\rho_{ij}}\]
with
\[M_{ij}^2=g_a g_b (1-\frac{n_i}{N_a^{val}})(1-\frac{n_j}{N_b^{val}})\] 
with $n_i$ and $n_j$ the occupation number of both lone-pairs, and:
\[\rho_{ij}=\frac{r_{ab}}{4\sqrt{r_{vdw_i}r_{vdw_j}}}\]
\[O_{1_{i,j}}=O_{i,j}S_{ij}\]
\[O_{2_{i,j}}=O_{i,j}S_{ij_2}\]
so that finally:
\[E_{i,j,rep-lp-lp}=n_{i}n_{j}(C_1\frac{O_{1_{i,j}}^2}{r_{ij}}+C_2\frac{O_{2_{i,j}}^2}{r_{ij}^2})\]
$n_i$ and $n_j$ the occupation numbers of the lone-pairs $i$ and $j$.
The cross bond-lone pair is defined in a similar manner:
\[E_{rep,bond-lp}=\sum_{i=1}^{nbond}\sum_{j=1}^{nlp}E_{i,j,bond-lp}\]
Let us denote by $a$ and $b$ the atoms forming the bond $i$ and by $c$ the atom carrying the lone pair $j$. We then have the two overlap terms:
\[O_{aj}=O_{s_a s_j}+O_{s_a p_j}+O_{s_j p_a}+O_{p_a p_j}\]
\[O_{bj}=O_{s_b s_j}+O_{s_b p_j}+O_{s_j p_b}+O_{p_b p_j}\]
Then, we have for example:
\[O_{s_a s_j}=C_{s_a}C_{s_j}\]
\[O_{s_a p_j}=C_{s_a}C_{p_j}f_{ac}cos(\vv{CA},\vv{CI})\]
\[O_{p_a p_j}=C_{p_a}C_{p_j}f_{ac}cos(\vv{AB},\vv{AC})cos(\vv{CA},\vv{CI})\]

\[S_{aj}=M_{aj}e^{-\alpha\rho_{aj}}\]
\[S_{aj_2}=M_{aj}e^{-\alpha_2\rho_{aj}}\] with
\[M_{aj}^2=g_a g_k (1-\frac{Q_a}{N_a^{val}})(1-\frac{n_j}{N_c^{val}})\] 
\[\rho_{aj}=\frac{r_{ak}}{4\sqrt{r_{vdw_a}r_{vdw_j}}}\]
\[O_{1_{i,j}}=O_{aj}S_{aj}+O_{bj}S_{bj}\]
\[O_{2_{i,j}}=O_{aj}S_{aj_2}+O_{bj}S_{bj_2}\]
Then if we define call $\vv{r_{mid_{ab}}}$ the position vector of the midpoint of $a$ and $b$ and by $\vv{r}=\vv{r_{mid_{ab}}}-\vv{r_j}$
\[E_{i,j,rep-bond-lp}=Occ_{ab}n_{j}(C_1\frac{O_{1_{i,j}}^2}{r}+C_2\frac{O_{2_{i,j}}^2}{r^2})\]
$n_j$ being the occupation number the lone pair $j$.
\subsection{Polarization}
The resolution of the polarization equations is similar to the one of the AMOEBA force field \cite{AMOEBA03}: let us consider a system of $N$ atoms, each bearing a multipole expansion (truncated at the quadrupole level) as permanent charge density and a polarizability tensor $\alpha_i$. 
 $\bE$  is the $3N$ vector gathering all electric fields $\vec{E_i}$ created by the permanent charge density at atomic position $i$, and $\bmu$ is the equivalent $3N$ vector gathering the induced dipoles experienced at each atomic site.
 $\bT$ is the $3N\times 3N$ polarization matrix, defined by block as follows. It bears the $3\times 3$ polarizability tensors $\alpha_i$ along its diagonal block, and the interaction between the $i$th and $j$th dipole is represented as the $T_{ij}$ tensor. 
     \[ \bT = \left(
     \begin{array}{ccccc}
      \alpha_1^{-1} & -T_{12} & -T_{13} & \ldots & -T_{1N}      \\
      -T_{21} & \alpha_2^{-1} & -T_{23} & \ldots & -T_{2N}      \\
      -T_{31} & -T_{32}       & \ddots  &        &              \\
      \vdots  & \vdots        &         &        &\vdots        \\
      -T_{N1} & -T_{N2}       &         & \ldots & \alpha_N^{-1}\\
     \end{array}
     \right)
     \]
     
This matrix is symmetric positive definite and such property is ensured by using a Thole damping function of the electric field at short-range to prevent any polarization catastrophe.\cite{scalablelip,SPMELagardere}

    Using these notations, the total polarization energy can be expressed as follows : 
    
  \begin{equation}
      E_{\text{pol}} = \frac12 \bmu^T \bT \bmu - \bmu^T \bE \label{eq:Epol}
  \end{equation}
    where $\bmu^T \bE$ represents the scalar product of vectors $\bmu$ and $\bE$ (also noted $\langle \bmu, \bE \rangle$).
	One can easily see that the dipole vector $\bmu$ minimizing (\ref{eq:Epol}) verifies the following linear system:
    \begin{equation}
    	\bT \bmu = \bE
    \end{equation}
	giving the minimized polarization energy: 
	\begin{equation}
    	E_{\text{pol}}  = - \frac12 \bmu^T \bE 
    \end{equation}
\subsection{Initial formulation of the Many-Body Charge Transfer}
The charge transfer formula and a thorough description of its computational complexity is described in the main text.

\subsection{Dispersion and Exchange-Dispersion}
\[E_{disp}=E_{disp-at-at}+E_{disp-lp-lp}+E_{disp-at-lp}\]
each of these terms corresponding to pairwise interactions, such that:
\[E_{disp-at-at}=\sum_{i,j=1,i < j}^{natoms}E_{i,j,at-at}\]
Let us define the reduced distance:
\[m_{ij}=\frac{r_{ij}}{2\sqrt{r_{vdw_i}r_{vdw_j}}}\]
with $r_{vdw_i}$ and $r_{vdw_j}$ van der Waals radii associated to atoms i and j. Let us also define the ratio:
\[d_{ij}=\frac{(r_{vdw_i}+r_{vdw_j})\alpha}{r_{ij}}-1\]
$\alpha$ being a global parameter and the pairwise parameter:
\[\beta_{ij}=\beta_i\beta_j\]
$\beta_i$ and $\beta_j$ being constants specific to atoms $i$ and $j$, then:
\[E_{i,j,at-at}=\beta_{ij}(C_6\frac{e^{-\gamma_6 d_{ij}}}{m_{ij}^6}+C_8\frac{e^{-\gamma_8 d_{ij}}}{m_{ij}^8}+C_{10}\frac{e^{-\gamma_{10} d_{ij}}}{m_{ij}^{10}})\]
with $\gamma_6$, $\gamma_8$ and $\gamma_{10}$, $C_6$, $C_8$ and $C_{10}$ global parameters.
Similarly, we have:
\[E_{disp-lp-lp}=\sum_{k,l=1,k < l}^{nlp}E_{k,l,lp-lp}\]
\[m_{kl}=\frac{r_{kl}}{2\sqrt{r_{vdw_k}r_{vdw_l}}}\]
with $r_{vdw_k}$ and $r_{vdw_l}$ van der Waals radii associated to lone-pairs k and l
\[d_{kl}=\frac{(r_{vdw_k}+r_{vdw_l})\alpha}{r_{kl}}-1\]
Then, denoting by $Q_k$ and $Q_l$ the charge of the atoms bearing respectively lone pair $k$ and lone pair $l$:
\[E_{k,l,lp-lp}=-\frac{Q_kQ_l}{4}(C_6\frac{e^{-\gamma_6 d_{kl}}}{m_{kl}^6}+C_8\frac{e^{-\gamma_8 d_{kl}}}{m_{kl}^8}+C_{10}\frac{e^{-\gamma_{10} d_{kl}}}{m_{kl}^{10}})\]
Finally, the cross atom-lone pair term reads:
\[E_{disp-atom-lp}=\sum_{i=1}^{natoms}\sum_{k=1}^{nlp}E_{i,k,atom-lp}\]
Let us denote by $c(k)$ the index of the atom carrying the lone pair k.
\[m_{ik}=\frac{r_{ik}}{2\sqrt{r_{vdw_i}r_{vdw_k}}}\]
\[d_{ik}=\frac{(r_{vdw_i}+r_{vdw_k})\alpha}{r_{ik}}-1\]
\[E_{i,k,atom-lp}=-\frac{\delta}{2}(C_6\frac{e^{-\gamma_6 d_{ik}}}{m_{ik}^6}+C_8\frac{e^{-\gamma_8 d_{ik}}}{m_{ik}^8}+C_{10}\frac{e^{-\gamma_{10} d_{ik}}}{m_{ik}^{10}})\]
$\delta$ being an additional scaling factor specific to atom-lone pair dispersion.

These three types of pairwise interactions are complemented by exponential-like short-range exchange dispersion terms, such that we can define:
\[E_{exchange-disp}=E_{xdisp-at-at}+E_{xdisp-lp-lp}+E_{xdisp-at-lp}\]
each of these terms also corresponding to pairwise interactions, such that:
\[E_{xdisp-at-at}=\sum_{i,j=1,i < j}^{natoms}Ex_{i,j,at-at}\]
Let us note $Q_i$ and $Q_j$ the partial charge carried by atom $i$ and $j$, $N_i^{val}$, and $N_j^{val}$ their number of valence electrons. 
Let us define:
\[K_{ij}=\beta_{ij}(1-\frac{Q_i}{N_i^{val}})(1-\frac{Q_j}{N_j^{val}})\]
$\beta_{ij}$ being defined before. Then,
\[Ex_{i,j,at-at}=C_1 K_{ij}e^{-A_1 m_{ij}}\]
with $m_{ij}$ defined as before and $C_1$, $A_1$ parameters specific to atom-atom exchange dispersion. Similarly:
\[E_{xdisp-lp-lp}=\sum_{k,l=1,k < l}^{nlp}Ex_{k,l,lp-lp}\]
\[K_{kl}=\beta_{c(k)c(l)}(1-\frac{Q_k}{N_{c(k)}^{val}})(1-\frac{Q_l}{N_{c(l)}^{val}})\]
\[Ex_{k,l,lp-lp}=\frac{n_k n_l}{4}C_2 K_{kl}e^{-A_2 m_{kl}}\]
with $n_k$ and $n_l$ and $m_{kl}$ defined as before and $C_2$, $A_2$ parameters specific to lone-pair lone-pair exchange dispersion.
\[E_{xdisp-at-lp}=\sum_{i=1}^{natoms}\sum_{k=1}^{nlp}Ex_{i,k,at-lp}\]
\[K_{ik}=\beta_{ic(k)}(1-\frac{Q_i}{N_i^{val}})(1-\frac{Q_k}{N_{c(k)}^{val}})\]
\[Ex_{i,k,at-lp}=\frac{n_k}{2}C_3 K_{ik}e^{-A_3 m_{ik}}\]
with $m_{ik}$ defined as before and $C_3$, $A_3$ parameters specific to atom lone-pair exchange dispersion.

\captionsetup{justification=justified,singlelinecheck=false}

\captionsetup[table]{singlelinecheck=false} 
\begin{table}[H]
\resizebox{\linewidth}{!}{
\begin{tabular}{lrrrrrrrrrrr}
\toprule
\midrule
Smith Geometries & \thead{1} & \thead{2} & \thead{3} & \thead{4} & \thead{5} & \thead{6} & \thead{7} & \thead{8} & \thead{9} & \thead{10} & \thead{RMSE} \\
\midrule
\cellcolor{mysilver}\textbf{Electrostatics} \\
\saptdft & -7.94 & -6.90 & -6.71 & -6.63 & -5.95 & -5.74 & -4.69 & -1.27 & -4.46 & -2.45 \\
\sibfa  & -7.57 & -6.49 & -6.33 & -6.55 & -6.17 & -6.14 & -4.28 & -1.06 & -3.84 & -2.19 & 0.36 \\
\\
\cellcolor{mysilver}\textbf{Exchange-Repulsion} \\
\saptdft  &  7.99 &  6.96 &  6.69 &  6.03 &  5.46 &  5.41 &  4.14 &  1.18 &  4.25 &  2.19 \\
\sibfa   &  6.92 &  6.03 &  6.11 &  6.41 &  6.06 &  7.04 &  3.60 &  0.63 &  3.29 &  1.68 & 0.85 \\
\\
\cellcolor{mysilver}\textbf{Polarization} \\
\saptdft  & -0.99 & -0.91 & -0.90 & -0.62 & -0.62 & -0.65 & -0.30 & -0.09 & -0.34 & -0.24 \\
\sibfa   & -1.11 & -1.04 & -1.03 & -0.67 & -0.69 & -0.72 & -0.28 & -0.07 & -0.42 & -0.28 & 0.08 \\
\\
\cellcolor{mysilver}\textbf{Charge Delocalization} \\
\saptdft   & -1.25 & -1.05 & -1.01 & -0.49 & -0.45 & -0.46 & -0.21 & -0.03 & -0.25 & -0.11 \\
\sibfa    & -1.06 & -0.93 & -0.95 & -0.42 & -0.36 & -0.40 & -0.19 & -0.04 & -0.30 & -0.16 & 0.09 \\
\\
\cellcolor{mysilver}\textbf{Dispersion} \\
\saptdft & -2.35 & -2.19 & -2.15 & -2.26 & -2.24 & -2.30 & -1.80 & -0.84 & -1.73 & -1.19 \\
\sibfa  & -2.14 & -2.05 & -2.05 & -2.21 & -2.27 & -2.41 & -1.72 & -0.76 & -1.58 & -1.15 & 0.11 \\
\\
\cellcolor{mysilver}\textbf{Binding Energy} \\
\ccsdt    & -4.80 & -4.27 & -4.25 & -4.20 & -3.98 & -3.89 & -2.96 & -1.03 & -2.56 & -1.75 \\
\saptdft  & -4.53 & -4.09 & -4.09 & -3.96 & -3.80 & -3.73 & -2.86 & -1.05 & -2.52 & -1.80 & 0.16 \\
\sibfa  & -4.98 & -4.48 & -4.24 & -3.45 & -3.44 & -2.63 & -2.88 & -1.30 & -2.87 & -2.11 & 0.53 \\
\bottomrule
\multicolumn{1}{@{} l}
{
\footnotesize{
\textsuperscript{\emph{a}} \EctREG{2}+\DELTAHF
}
}
\end{tabular}
}
\caption{Comparison of \sibfa and \saptdft energy components for the Smith water dimers. The PBE0 functional in \saptdft, and the aug-cc-pVTZ basis were used in both \saptdft and \ccsdt calculations.
}
\label{tab-sibfa-vs-saptdft-smith-dimers}
\end{table}

\begin{table}[H]
\centering
\resizebox{!}{.35\paperheight}{%
\begin{tabular}{lcc}
\toprule 
\midrule
Clusters & Ref & \sibfa  \\
\midrule
\cellcolor{mysilver}{Smith Dimers}  \\
1	&	-4.97(-4.80)		&	-4.98		\\
2	&	-4.45(-4.27)		&	-4.48		\\
3	&	-4.42(-4.25)		&	-4.24		\\
4	&	-4.25(-4.20)		&	-3.45		\\
5	&	-4.00(-3.98)		&	-3.44		\\
6	&	-3.96(-3.89)	    &	-2.63		\\
7	&	-3.26(-2.96)		&	-2.88		\\
8	&	-1.30(-1.03)		&	-1.30		\\
9	&	-3.05(-2.56)		&	-2.87		\\
10	&	-2.18(-1.75)		&	-2.11		\\

\midrule
trimer	&	-15.74	&	-16.20		\\
tretramer	&	-27.40	&	-27.97	\\
pentamer	&	-35.93	&	-35.91		\\
\cellcolor{mysilver}{Hexamers}  \\											
prism	&	-45.92	&	-47.48		\\
cage	&	-45.67	&	-47.11		\\
bag	&	-44.30	&	-45.05			\\
cyclic chair	&	-44.12 &	-43.62		\\
book1	&	-45.20	&	-45.78		\\
book2	&	-44.90	&	-46.27		\\
cyclic boat1	&	-43.13	&	-42.80		\\
cyclic boat2	&	-43.07	&	-43.12		\\
\cellcolor{mysilver}{Octamers} \\											
S$_4$	&	-72.70	&	-76.67		\\
D$_{2d}$	&	-72.70	&	-76.79		\\
\cellcolor{mysilver}{Endecamers} \\											
434	&	-105.72	&	-106.35		\\
515	&	-105.18	&	-104.37		\\
551	&	-104.92	&	-104.34		\\
443	&	-104.76	&	-106.64		\\
4412	&	-103.97	&	-104.40		\\
\cellcolor{mysilver}{16 mers} \\											
boat-a	&	-170.80	&	-168.15		\\
boat-b	&	-170.63	&	-168.61		\\
antiboat	&	-170.54	&	-167.57		\\
ABAB	&	-171.05	&	-173.25		\\
AABB	&	-170.51	&	-171.86		\\
\cellcolor{mysilver}{17 mers} \\											
sphere	&	-182.54	&	-179.10		\\
5525	&	-181.83	&	-178.38		\\
\cellcolor{mysilver}{20 mers} \\											
dodecahedron	&	-200.10	&	-200.67		\\
fused cubes	&	-212.10	&	-221.22		\\
face sharing prisms	&	-215.20	&	-217.53		\\
edge sharing prisms	&	-218.10	&	-219.28		\\

\midrule
\bottomrule
\end{tabular}%
}
\caption{Comparison of the binding energies computed with \sibfa, to the \textit{ab initio} reference computed at different level of theories : \ccsdt/CBS (or \ccsdt/aug-cc-PVTZ in parenthesis) for the Smith dimers \cite{Smith1990}, trimer, tetramer, pentamer  \cite{ErinE.Dahlke2008,Bates2009} and hexamers \cite{Wang2013}, MP2/CBS for octamers \cite{Xantheas2004} and 20 mers \cite{Fanourgakis2004}, MP2/aug-cc-pV5Z (434,515,551) and MP2/aug-cc-pVQZ (443,4412) for endecamers \cite{Bulusu2006} and \ccsdt/aug-cc-pVTZ for 16 and 17 mers \cite{Yoo2010}. Values in \kcalmol.}
\label{tab-ebind}
\end{table}

\captionsetup{justification=justified,singlelinecheck=false}

\captionsetup{justification=justified,singlelinecheck=false}

\begin{acknowledgement}
This work has received funding from the European Research Council (ERC) under the European Union’s Horizon 2020 research and innovation program (grant agreement No 810367), project EMC2 (JPP). Computations have been performed at GENCI (IDRIS, Orsay, France and TGCC, Bruyères le Chatel) on grant no A0070707671. PR and GAC are grateful for support by National Institutes of Health (PR: R01GM106137 and R01GM114237; GAC:R01GM108583). PR acknowledges funding from the Cancer Prevention and National Science Foundation (CHE-1856173). JPP thank Alston Misquitta (Queen Mary, London, UK) for providing the latest development version of the CamCasp code that was used for the SAPT(DFT) computations. JPP and NG thank also Jay W. Ponder (Washington University in Saint Louis, USA) for discussions since the inception of the project.
\end{acknowledgement}

\begin{suppinfo}

(1) Energy tables for the linear water dimers. (2) Energy table for the water clusters. (3) Auxiliary Figures and Tables. (4) Equations used to compute the condensed phase properties. (5) Key, LP and parameters files for \sibfa MD with \tinkerhp.

\end{suppinfo}

\newpage

\bibliography{biblio.bib}

\providecommand{\latin}[1]{#1}
\makeatletter
\providecommand{\doi}
  {\begingroup\let\do\@makeother\dospecials
  \catcode`\{=1 \catcode`\}=2 \doi@aux}
\providecommand{\doi@aux}[1]{\endgroup\texttt{#1}}
\makeatother
\providecommand*\mcitethebibliography{\thebibliography}
\csname @ifundefined\endcsname{endmcitethebibliography}
  {\let\endmcitethebibliography\endthebibliography}{}
\begin{mcitethebibliography}{100}
\providecommand*\natexlab[1]{#1}
\providecommand*\mciteSetBstSublistMode[1]{}
\providecommand*\mciteSetBstMaxWidthForm[2]{}
\providecommand*\mciteBstWouldAddEndPuncttrue
  {\def\EndOfBibitem{\unskip.}}
\providecommand*\mciteBstWouldAddEndPunctfalse
  {\let\EndOfBibitem\relax}
\providecommand*\mciteSetBstMidEndSepPunct[3]{}
\providecommand*\mciteSetBstSublistLabelBeginEnd[3]{}
\providecommand*\EndOfBibitem{}
\mciteSetBstSublistMode{f}
\mciteSetBstMaxWidthForm{subitem}{(\alph{mcitesubitemcount})}
\mciteSetBstSublistLabelBeginEnd
  {\mcitemaxwidthsubitemform\space}
  {\relax}
  {\relax}

\bibitem[Stone(2013)]{stone2013theory}
Stone,~A. \emph{The theory of intermolecular forces}; Oxford University Press,
  Oxford, UK., 2013\relax
\mciteBstWouldAddEndPuncttrue
\mciteSetBstMidEndSepPunct{\mcitedefaultmidpunct}
{\mcitedefaultendpunct}{\mcitedefaultseppunct}\relax
\EndOfBibitem
\bibitem[Gresh \latin{et~al.}(2007)Gresh, Cisneros, Darden, and
  Piquemal]{NohadGresh2007}
Gresh,~N.; Cisneros,~G.~A.; Darden,~T.~A.; Piquemal,~J.-P. {Anisotropic,
  Polarizable Molecular Mechanics Studies of Inter- and Intramolecular
  Interactions and Ligand-Macromolecule Complexes. A Bottom-Up Strategy}.
  \emph{J. Chem. Theory Comput.} \textbf{2007}, \emph{3}, 1960--1986\relax
\mciteBstWouldAddEndPuncttrue
\mciteSetBstMidEndSepPunct{\mcitedefaultmidpunct}
{\mcitedefaultendpunct}{\mcitedefaultseppunct}\relax
\EndOfBibitem
\bibitem[Warshel \latin{et~al.}(2007)Warshel, Kato, and Pisliakov]{warshelrev}
Warshel,~A.; Kato,~M.; Pisliakov,~A.~V. Polarizable Force Fields: History, Test
  Cases, and Prospects. \emph{Journal of Chemical Theory and Computation}
  \textbf{2007}, \emph{3}, 2034--2045, PMID: 26636199\relax
\mciteBstWouldAddEndPuncttrue
\mciteSetBstMidEndSepPunct{\mcitedefaultmidpunct}
{\mcitedefaultendpunct}{\mcitedefaultseppunct}\relax
\EndOfBibitem
\bibitem[Shi \latin{et~al.}(2015)Shi, Ren, Schnieders, and
  Piquemal]{doi:https://doi.org/10.1002/9781118889886.ch2}
Shi,~Y.; Ren,~P.; Schnieders,~M.; Piquemal,~J.-P. \emph{Reviews in
  Computational Chemistry Volume 28}; John Wiley \& Sons, Ltd, 2015; Chapter 2,
  pp 51--86\relax
\mciteBstWouldAddEndPuncttrue
\mciteSetBstMidEndSepPunct{\mcitedefaultmidpunct}
{\mcitedefaultendpunct}{\mcitedefaultseppunct}\relax
\EndOfBibitem
\bibitem[Melcr and Piquemal(2019)Melcr, and Piquemal]{10.3389/fmolb.2019.00143}
Melcr,~J.; Piquemal,~J.-P. Accurate Biomolecular Simulations Account for
  Electronic Polarization. \emph{Frontiers in Molecular Biosciences}
  \textbf{2019}, \emph{6}, 143\relax
\mciteBstWouldAddEndPuncttrue
\mciteSetBstMidEndSepPunct{\mcitedefaultmidpunct}
{\mcitedefaultendpunct}{\mcitedefaultseppunct}\relax
\EndOfBibitem
\bibitem[Jing \latin{et~al.}(2019)Jing, Liu, Cheng, Qi, Walker, Piquemal, and
  Ren]{doi:10.1146/annurev-biophys-070317-033349}
Jing,~Z.; Liu,~C.; Cheng,~S.~Y.; Qi,~R.; Walker,~B.~D.; Piquemal,~J.-P.;
  Ren,~P. Polarizable Force Fields for Biomolecular Simulations: Recent
  Advances and Applications. \emph{Annual Review of Biophysics} \textbf{2019},
  \emph{48}, 371--394, PMID: 30916997\relax
\mciteBstWouldAddEndPuncttrue
\mciteSetBstMidEndSepPunct{\mcitedefaultmidpunct}
{\mcitedefaultendpunct}{\mcitedefaultseppunct}\relax
\EndOfBibitem
\bibitem[Bedrov \latin{et~al.}(2019)Bedrov, Piquemal, Borodin, MacKerell, Roux,
  and Schröder]{doi:10.1021/acs.chemrev.8b00763}
Bedrov,~D.; Piquemal,~J.-P.; Borodin,~O.; MacKerell,~A.~D.; Roux,~B.;
  Schröder,~C. Molecular Dynamics Simulations of Ionic Liquids and
  Electrolytes Using Polarizable Force Fields. \emph{Chemical Reviews}
  \textbf{2019}, \emph{119}, 7940--7995, PMID: 31141351\relax
\mciteBstWouldAddEndPuncttrue
\mciteSetBstMidEndSepPunct{\mcitedefaultmidpunct}
{\mcitedefaultendpunct}{\mcitedefaultseppunct}\relax
\EndOfBibitem
\bibitem[Ren and Ponder(2003)Ren, and Ponder]{AMOEBA03}
Ren,~P.; Ponder,~J.~W. Polarizable Atomic Multipole Water Model for Molecular
  Mechanics Simulation. \emph{The Journal of Physical Chemistry B}
  \textbf{2003}, \emph{107}, 5933--5947\relax
\mciteBstWouldAddEndPuncttrue
\mciteSetBstMidEndSepPunct{\mcitedefaultmidpunct}
{\mcitedefaultendpunct}{\mcitedefaultseppunct}\relax
\EndOfBibitem
\bibitem[Lamoureux \latin{et~al.}(2003)Lamoureux, MacKerell, and
  Roux]{Drudewater}
Lamoureux,~G.; MacKerell,~A.~D.; Roux,~B. A simple polarizable model of water
  based on classical Drude oscillators. \emph{The Journal of Chemical Physics}
  \textbf{2003}, \emph{119}, 5185--5197\relax
\mciteBstWouldAddEndPuncttrue
\mciteSetBstMidEndSepPunct{\mcitedefaultmidpunct}
{\mcitedefaultendpunct}{\mcitedefaultseppunct}\relax
\EndOfBibitem
\bibitem[Fanourgakis and Xantheas(2008)Fanourgakis, and Xantheas]{TTM3-F}
Fanourgakis,~G.~S.; Xantheas,~S.~S. Development of transferable interaction
  potentials for water. V. Extension of the flexible, polarizable, Thole-type
  model potential (TTM3-F, v. 3.0) to describe the vibrational spectra of water
  clusters and liquid water. \emph{The Journal of Chemical Physics}
  \textbf{2008}, \emph{128}, 074506\relax
\mciteBstWouldAddEndPuncttrue
\mciteSetBstMidEndSepPunct{\mcitedefaultmidpunct}
{\mcitedefaultendpunct}{\mcitedefaultseppunct}\relax
\EndOfBibitem
\bibitem[Liu \latin{et~al.}(2019)Liu, Piquemal, and Ren]{Liu2019}
Liu,~C.; Piquemal,~J.-P.; Ren,~P. {AMOEBA+ Classical Potential for Modeling
  Molecular Interactions}. \emph{J. Chem. Theory Comput.} \textbf{2019},
  \emph{15}, 4122--4139\relax
\mciteBstWouldAddEndPuncttrue
\mciteSetBstMidEndSepPunct{\mcitedefaultmidpunct}
{\mcitedefaultendpunct}{\mcitedefaultseppunct}\relax
\EndOfBibitem
\bibitem[Liu \latin{et~al.}(2020)Liu, Piquemal, and Ren]{Liu2020}
Liu,~C.; Piquemal,~J.-P.; Ren,~P. {Implementation of Geometry-Dependent Charge
  Flux into the Polarizable AMOEBA+ Potential}. \emph{J. Phys. Chem. Lett}
  \textbf{2020}, \emph{2020}, 419--426\relax
\mciteBstWouldAddEndPuncttrue
\mciteSetBstMidEndSepPunct{\mcitedefaultmidpunct}
{\mcitedefaultendpunct}{\mcitedefaultseppunct}\relax
\EndOfBibitem
\bibitem[Babin \latin{et~al.}(2013)Babin, Leforestier, and Paesani]{MBPOL}
Babin,~V.; Leforestier,~C.; Paesani,~F. Development of a “First Principles”
  Water Potential with Flexible Monomers: Dimer Potential Energy Surface, VRT
  Spectrum, and Second Virial Coefficient. \emph{Journal of Chemical Theory and
  Computation} \textbf{2013}, \emph{9}, 5395--5403, PMID: 26592277\relax
\mciteBstWouldAddEndPuncttrue
\mciteSetBstMidEndSepPunct{\mcitedefaultmidpunct}
{\mcitedefaultendpunct}{\mcitedefaultseppunct}\relax
\EndOfBibitem
\bibitem[Naseem-Khan \latin{et~al.}(2021)Naseem-Khan, Piquemal, and
  Cisneros]{GEM}
Naseem-Khan,~S.; Piquemal,~J.-P.; Cisneros,~G.~A. Improvement of the Gaussian
  Electrostatic Model by separate fitting of Coulomb and exchange-repulsion
  densities and implementation of a new dispersion term. \emph{The Journal of
  Chemical Physics} \textbf{2021}, \emph{155}, 194103\relax
\mciteBstWouldAddEndPuncttrue
\mciteSetBstMidEndSepPunct{\mcitedefaultmidpunct}
{\mcitedefaultendpunct}{\mcitedefaultseppunct}\relax
\EndOfBibitem
\bibitem[Cisneros \latin{et~al.}(2016)Cisneros, Wikfeldt, Ojamäe, Lu, Xu,
  Torabifard, Bartók, Csányi, Molinero, and Paesani]{paesanireview}
Cisneros,~G.~A.; Wikfeldt,~K.~T.; Ojamäe,~L.; Lu,~J.; Xu,~Y.; Torabifard,~H.;
  Bartók,~A.~P.; Csányi,~G.; Molinero,~V.; Paesani,~F. Modeling Molecular
  Interactions in Water: From Pairwise to Many-Body Potential Energy Functions.
  \emph{Chemical Reviews} \textbf{2016}, \emph{116}, 7501--7528, PMID:
  27186804\relax
\mciteBstWouldAddEndPuncttrue
\mciteSetBstMidEndSepPunct{\mcitedefaultmidpunct}
{\mcitedefaultendpunct}{\mcitedefaultseppunct}\relax
\EndOfBibitem
\bibitem[Das \latin{et~al.}(2019)Das, Urban, Leven, Loipersberger, Aldossary,
  Head-Gordon, and Head-Gordon]{Das2019}
Das,~A.~K.; Urban,~L.; Leven,~I.; Loipersberger,~M.; Aldossary,~A.;
  Head-Gordon,~M.; Head-Gordon,~T. {Development of an Advanced Force Field for
  Water Using Variational Energy Decomposition Analysis}. \emph{J. Chem. Theory
  Comput.} \textbf{2019}, \emph{15}, 5001--5013\relax
\mciteBstWouldAddEndPuncttrue
\mciteSetBstMidEndSepPunct{\mcitedefaultmidpunct}
{\mcitedefaultendpunct}{\mcitedefaultseppunct}\relax
\EndOfBibitem
\bibitem[Khaliullin \latin{et~al.}(2007)Khaliullin, Cobar, Lochan, Bell, and
  Head-Gordon]{Khaliullin2007}
Khaliullin,~R.~Z.; Cobar,~E.~A.; Lochan,~R.~C.; Bell,~A.~T.; Head-Gordon,~M.
  {Unravelling the origin of intermolecular interactions using absolutely
  localized molecular orbitals}. \emph{J. Phys. Chem. A} \textbf{2007},
  \emph{111}, 8753--8765\relax
\mciteBstWouldAddEndPuncttrue
\mciteSetBstMidEndSepPunct{\mcitedefaultmidpunct}
{\mcitedefaultendpunct}{\mcitedefaultseppunct}\relax
\EndOfBibitem
\bibitem[Jeziorski \latin{et~al.}(1994)Jeziorski, Moszynski, and
  Szalewicz]{Jeziorski1994}
Jeziorski,~B.; Moszynski,~R.; Szalewicz,~K. {Perturbation Theory Approach to
  Intermolecular Potential Energy Surfaces of van der Waals Complexes.}
  \emph{Chem. Rev.} \textbf{1994}, \emph{94}, 1887--1930\relax
\mciteBstWouldAddEndPuncttrue
\mciteSetBstMidEndSepPunct{\mcitedefaultmidpunct}
{\mcitedefaultendpunct}{\mcitedefaultseppunct}\relax
\EndOfBibitem
\bibitem[Parker \latin{et~al.}(2014)Parker, Burns, Parrish, Ryno, and
  Sherrill]{Parker2014}
Parker,~T.~M.; Burns,~L.~A.; Parrish,~R.~M.; Ryno,~A.~G.; Sherrill,~C.~D.
  {Levels of symmetry adapted perturbation theory (SAPT). I. Efficiency and
  performance for interaction energies}. \emph{J. Chem. Phys} \textbf{2014},
  \emph{140}\relax
\mciteBstWouldAddEndPuncttrue
\mciteSetBstMidEndSepPunct{\mcitedefaultmidpunct}
{\mcitedefaultendpunct}{\mcitedefaultseppunct}\relax
\EndOfBibitem
\bibitem[Gordon \latin{et~al.}(2007)Gordon, Slipchenko, Li, and
  Jensen]{GORDON2007177}
Gordon,~M.~S.; Slipchenko,~L.; Li,~H.; Jensen,~J.~H. In \emph{Chapter 10 The
  Effective Fragment Potential: A General Method for Predicting Intermolecular
  Interactions}; Spellmeyer,~D., Wheeler,~R., Eds.; Annual Reports in
  Computational Chemistry; Elsevier, 2007; Vol.~3; pp 177--193\relax
\mciteBstWouldAddEndPuncttrue
\mciteSetBstMidEndSepPunct{\mcitedefaultmidpunct}
{\mcitedefaultendpunct}{\mcitedefaultseppunct}\relax
\EndOfBibitem
\bibitem[Gordon \latin{et~al.}(2013)Gordon, Smith, Xu, and Slipchenko]{EFP}
Gordon,~M.~S.; Smith,~Q.~A.; Xu,~P.; Slipchenko,~L.~V. Accurate First
  Principles Model Potentials for Intermolecular Interactions. \emph{Annual
  Review of Physical Chemistry} \textbf{2013}, \emph{64}, 553--578, PMID:
  23561011\relax
\mciteBstWouldAddEndPuncttrue
\mciteSetBstMidEndSepPunct{\mcitedefaultmidpunct}
{\mcitedefaultendpunct}{\mcitedefaultseppunct}\relax
\EndOfBibitem
\bibitem[Bukowski \latin{et~al.}(2007)Bukowski, Szalewicz, Groenenboom, and
  van~der Avoird]{SAPT5s}
Bukowski,~R.; Szalewicz,~K.; Groenenboom,~G.~C.; van~der Avoird,~A. Predictions
  of the Properties of Water from First Principles. \emph{Science}
  \textbf{2007}, \emph{315}, 1249--1252\relax
\mciteBstWouldAddEndPuncttrue
\mciteSetBstMidEndSepPunct{\mcitedefaultmidpunct}
{\mcitedefaultendpunct}{\mcitedefaultseppunct}\relax
\EndOfBibitem
\bibitem[Reddy \latin{et~al.}(2016)Reddy, Straight, Bajaj, {Huy Pham}, Riera,
  Moberg, Morales, Knight, G{\"{o}}tz, and Paesani]{Reddy2016}
Reddy,~S.~K.; Straight,~S.~C.; Bajaj,~P.; {Huy Pham},~C.; Riera,~M.;
  Moberg,~D.~R.; Morales,~M.~A.; Knight,~C.; G{\"{o}}tz,~A.~W.; Paesani,~F. {On
  the accuracy of the MB-pol many-body potential for water: Interaction
  energies, vibrational frequencies, and classical thermodynamic and dynamical
  properties from clusters to liquid water and ice}. \emph{J. Chem. Phys.}
  \textbf{2016}, \emph{145}\relax
\mciteBstWouldAddEndPuncttrue
\mciteSetBstMidEndSepPunct{\mcitedefaultmidpunct}
{\mcitedefaultendpunct}{\mcitedefaultseppunct}\relax
\EndOfBibitem
\bibitem[Kumar \latin{et~al.}(2010)Kumar, Wang, Jenness, and Jordan]{DPP}
Kumar,~R.; Wang,~F.-F.; Jenness,~G.~R.; Jordan,~K.~D. A second generation
  distributed point polarizable water model. \emph{The Journal of Chemical
  Physics} \textbf{2010}, \emph{132}, 014309\relax
\mciteBstWouldAddEndPuncttrue
\mciteSetBstMidEndSepPunct{\mcitedefaultmidpunct}
{\mcitedefaultendpunct}{\mcitedefaultseppunct}\relax
\EndOfBibitem
\bibitem[Piquemal \latin{et~al.}(2006)Piquemal, Cisneros, Reinhardt, Gresh, and
  Darden]{GEM0}
Piquemal,~J.-P.; Cisneros,~G.~A.; Reinhardt,~P.; Gresh,~N.; Darden,~T.~A.
  Towards a force field based on density fitting. \emph{The Journal of Chemical
  Physics} \textbf{2006}, \emph{124}, 104101\relax
\mciteBstWouldAddEndPuncttrue
\mciteSetBstMidEndSepPunct{\mcitedefaultmidpunct}
{\mcitedefaultendpunct}{\mcitedefaultseppunct}\relax
\EndOfBibitem
\bibitem[Duke \latin{et~al.}(2014)Duke, Starovoytov, Piquemal, and
  Cisneros]{Gemstar}
Duke,~R.~E.; Starovoytov,~O.~N.; Piquemal,~J.-P.; Cisneros,~G.~A. GEM*: A
  Molecular Electronic Density-Based Force Field for Molecular Dynamics
  Simulations. \emph{Journal of Chemical Theory and Computation} \textbf{2014},
  \emph{10}, 1361--1365, PMID: 26580355\relax
\mciteBstWouldAddEndPuncttrue
\mciteSetBstMidEndSepPunct{\mcitedefaultmidpunct}
{\mcitedefaultendpunct}{\mcitedefaultseppunct}\relax
\EndOfBibitem
\bibitem[Naseem-Khan \latin{et~al.}(2021)Naseem-Khan, Gresh, Misquitta, and
  Piquemal]{SAPTSehr}
Naseem-Khan,~S.; Gresh,~N.; Misquitta,~A.~J.; Piquemal,~J.-P. Assessment of
  SAPT and Supermolecular EDA Approaches for the Development of Separable and
  Polarizable Force Fields. \emph{Journal of Chemical Theory and Computation}
  \textbf{2021}, \emph{17}, 2759--2774, PMID: 33877844\relax
\mciteBstWouldAddEndPuncttrue
\mciteSetBstMidEndSepPunct{\mcitedefaultmidpunct}
{\mcitedefaultendpunct}{\mcitedefaultseppunct}\relax
\EndOfBibitem
\bibitem[Rackers \latin{et~al.}(2021)Rackers, Silva, Wang, and Ponder]{HIPPO}
Rackers,~J.~A.; Silva,~R.~R.; Wang,~Z.; Ponder,~J.~W. Polarizable Water
  Potential Derived from a Model Electron Density. \emph{Journal of Chemical
  Theory and Computation} \textbf{2021}, \emph{17}, 7056--7084, PMID:
  34699197\relax
\mciteBstWouldAddEndPuncttrue
\mciteSetBstMidEndSepPunct{\mcitedefaultmidpunct}
{\mcitedefaultendpunct}{\mcitedefaultseppunct}\relax
\EndOfBibitem
\bibitem[Gilmore \latin{et~al.}(2020)Gilmore, Dove, and Misquitta]{DIFF}
Gilmore,~R. A.~J.; Dove,~M.~T.; Misquitta,~A.~J. First-Principles Many-Body
  Nonadditive Polarization Energies from Monomer and Dimer Calculations Only: A
  Case Study on Water. \emph{Journal of Chemical Theory and Computation}
  \textbf{2020}, \emph{16}, 224--242, PMID: 31769980\relax
\mciteBstWouldAddEndPuncttrue
\mciteSetBstMidEndSepPunct{\mcitedefaultmidpunct}
{\mcitedefaultendpunct}{\mcitedefaultseppunct}\relax
\EndOfBibitem
\bibitem[Gresh \latin{et~al.}(1982)Gresh, Claverie, and Pullman]{Gresh1982}
Gresh,~N.; Claverie,~P.; Pullman,~A. {Computations of intermolecular
  interactions: Expansion of a charge‐transfer energy contribution in the
  framework of an additive procedure. Applications to hydrogen‐bonded
  systems}. \emph{Int. J. Quantum Chem.} \textbf{1982}, \emph{22},
  199--215\relax
\mciteBstWouldAddEndPuncttrue
\mciteSetBstMidEndSepPunct{\mcitedefaultmidpunct}
{\mcitedefaultendpunct}{\mcitedefaultseppunct}\relax
\EndOfBibitem
\bibitem[Gresh \latin{et~al.}(1986)Gresh, Claverie, and Pullman]{Gresh1986}
Gresh,~N.; Claverie,~P.; Pullman,~A. Intermolecular interactions: Elaboration
  on an additive procedure including an explicit charge-transfer contribution.
  \emph{International Journal of Quantum Chemistry} \textbf{1986}, \emph{29},
  101--118\relax
\mciteBstWouldAddEndPuncttrue
\mciteSetBstMidEndSepPunct{\mcitedefaultmidpunct}
{\mcitedefaultendpunct}{\mcitedefaultseppunct}\relax
\EndOfBibitem
\bibitem[Gresh(1995)]{Gresh1995}
Gresh,~N. {Energetics of Zn2+ binding to a series of biologically relevant
  ligands: A molecular mechanics investigation grounded on ab initio
  self‐consistent field supermolecular computations}. \emph{J. Comput. Chem.}
  \textbf{1995}, \emph{16}, 856--882\relax
\mciteBstWouldAddEndPuncttrue
\mciteSetBstMidEndSepPunct{\mcitedefaultmidpunct}
{\mcitedefaultendpunct}{\mcitedefaultseppunct}\relax
\EndOfBibitem
\bibitem[Piquemal \latin{et~al.}(2007)Piquemal, Chevreau, and
  Gresh]{JPPseparability2007}
Piquemal,~J.-P.; Chevreau,~H.; Gresh,~N. Toward a Separate Reproduction of the
  Contributions to the Hartree Fock and DFT Intermolecular Interaction Energies
  by Polarizable Molecular Mechanics with the SIBFA Potential. \emph{Journal of
  Chemical Theory and Computation} \textbf{2007}, \emph{3}, 824--837, PMID:
  26627402\relax
\mciteBstWouldAddEndPuncttrue
\mciteSetBstMidEndSepPunct{\mcitedefaultmidpunct}
{\mcitedefaultendpunct}{\mcitedefaultseppunct}\relax
\EndOfBibitem
\bibitem[Piquemal \latin{et~al.}(2005)Piquemal, Marquez, Parisel, and
  Giessner–Prettre]{CSOV}
Piquemal,~J.-P.; Marquez,~A.; Parisel,~O.; Giessner–Prettre,~C. A CSOV study
  of the difference between HF and DFT intermolecular interaction energy
  values: The importance of the charge transfer contribution. \emph{Journal of
  Computational Chemistry} \textbf{2005}, \emph{26}, 1052--1062\relax
\mciteBstWouldAddEndPuncttrue
\mciteSetBstMidEndSepPunct{\mcitedefaultmidpunct}
{\mcitedefaultendpunct}{\mcitedefaultseppunct}\relax
\EndOfBibitem
\bibitem[Ren and Ponder(2003)Ren, and Ponder]{Ren2003}
Ren,~P.; Ponder,~J.~W. {Polarizable Atomic Multipole Water Model for Molecular
  Mechanics Simulation}. \emph{J. Phys. Chem. B} \textbf{2003}, \emph{107},
  5933--5947\relax
\mciteBstWouldAddEndPuncttrue
\mciteSetBstMidEndSepPunct{\mcitedefaultmidpunct}
{\mcitedefaultendpunct}{\mcitedefaultseppunct}\relax
\EndOfBibitem
\bibitem[Laury \latin{et~al.}(2015)Laury, Wang, Pande, Head-Gordon, and
  Ponder]{Laury2015}
Laury,~M.~L.; Wang,~L.~P.; Pande,~V.~S.; Head-Gordon,~T.; Ponder,~J.~W.
  {Revised Parameters for the AMOEBA Polarizable Atomic Multipole Water Model}.
  \emph{J. Phys. Chem. B} \textbf{2015}, \emph{119}, 9423--9437\relax
\mciteBstWouldAddEndPuncttrue
\mciteSetBstMidEndSepPunct{\mcitedefaultmidpunct}
{\mcitedefaultendpunct}{\mcitedefaultseppunct}\relax
\EndOfBibitem
\bibitem[Lagard{\`{e}}re \latin{et~al.}(2018)Lagard{\`{e}}re, Jolly, Lipparini,
  Aviat, Stamm, Jing, Harger, Torabifard, Cisneros, Schnieders, Gresh, Maday,
  Ren, Ponder, and Piquemal]{Lagardere2018}
Lagard{\`{e}}re,~L.; Jolly,~L.-H.; Lipparini,~F.; Aviat,~F.; Stamm,~B.;
  Jing,~Z.~F.; Harger,~M.; Torabifard,~H.; Cisneros,~G.~A.; Schnieders,~M.~J.;
  Gresh,~N.; Maday,~Y.; Ren,~P.~Y.; Ponder,~J.~W.; Piquemal,~J.-P. {Tinker-HP:
  a massively parallel molecular dynamics package for multiscale simulations of
  large complex systems with advanced point dipole polarizable force fields}.
  \emph{Chem. Sci.} \textbf{2018}, \emph{9}, 956--972\relax
\mciteBstWouldAddEndPuncttrue
\mciteSetBstMidEndSepPunct{\mcitedefaultmidpunct}
{\mcitedefaultendpunct}{\mcitedefaultseppunct}\relax
\EndOfBibitem
\bibitem[Essmann \latin{et~al.}(1995)Essmann, Perera, Berkowitz, Darden, Lee,
  and Pedersen]{SPME}
Essmann,~U.; Perera,~L.; Berkowitz,~M.~L.; Darden,~T.; Lee,~H.; Pedersen,~L.~G.
  A smooth particle mesh Ewald method. \emph{The Journal of Chemical Physics}
  \textbf{1995}, \emph{103}, 8577--8593\relax
\mciteBstWouldAddEndPuncttrue
\mciteSetBstMidEndSepPunct{\mcitedefaultmidpunct}
{\mcitedefaultendpunct}{\mcitedefaultseppunct}\relax
\EndOfBibitem
\bibitem[Lagardère \latin{et~al.}(2015)Lagardère, Lipparini, Polack, Stamm,
  Cancès, Schnieders, Ren, Maday, and Piquemal]{SPMELagardere}
Lagardère,~L.; Lipparini,~F.; Polack,~E.; Stamm,~B.; Cancès,~E.;
  Schnieders,~M.; Ren,~P.; Maday,~Y.; Piquemal,~J.-P. Scalable Evaluation of
  Polarization Energy and Associated Forces in Polarizable Molecular Dynamics:
  II. Toward Massively Parallel Computations Using Smooth Particle Mesh Ewald.
  \emph{Journal of Chemical Theory and Computation} \textbf{2015}, \emph{11},
  2589--2599, PMID: 26575557\relax
\mciteBstWouldAddEndPuncttrue
\mciteSetBstMidEndSepPunct{\mcitedefaultmidpunct}
{\mcitedefaultendpunct}{\mcitedefaultseppunct}\relax
\EndOfBibitem
\bibitem[Misquitta \latin{et~al.}(2005)Misquitta, Podeszwa, Jeziorski, and
  Szalewicz]{misquitta_saptdft_2005}
Misquitta,~A.~J.; Podeszwa,~R.; Jeziorski,~B.; Szalewicz,~K. Intermolecular
  Potentials Based on Symmetry-Adapted Perturbation Theory with Dispersion
  Energies from Time-Dependent Density-Fun ctional Calculations. \emph{J. Chem.
  Phys.} \textbf{2005}, \emph{123}, 214103\relax
\mciteBstWouldAddEndPuncttrue
\mciteSetBstMidEndSepPunct{\mcitedefaultmidpunct}
{\mcitedefaultendpunct}{\mcitedefaultseppunct}\relax
\EndOfBibitem
\bibitem[Allinger \latin{et~al.}(1989)Allinger, Yuh, and Lii]{MM3}
Allinger,~N.~L.; Yuh,~Y.~H.; Lii,~J.~H. Molecular mechanics. The MM3 force
  field for hydrocarbons. 1. \emph{Journal of the American Chemical Society}
  \textbf{1989}, \emph{111}, 8551--8566\relax
\mciteBstWouldAddEndPuncttrue
\mciteSetBstMidEndSepPunct{\mcitedefaultmidpunct}
{\mcitedefaultendpunct}{\mcitedefaultseppunct}\relax
\EndOfBibitem
\bibitem[Urey and Bradley~Jr(1931)Urey, and Bradley~Jr]{urey1931vibrations}
Urey,~H.~C.; Bradley~Jr,~C.~A. The vibrations of pentatonic tetrahedral
  molecules. \emph{Physical review} \textbf{1931}, \emph{38}, 1969\relax
\mciteBstWouldAddEndPuncttrue
\mciteSetBstMidEndSepPunct{\mcitedefaultmidpunct}
{\mcitedefaultendpunct}{\mcitedefaultseppunct}\relax
\EndOfBibitem
\bibitem[Cisneros \latin{et~al.}(2006)Cisneros, Piquemal, and Darden]{GGEM}
Cisneros,~G.~A.; Piquemal,~J.-P.; Darden,~T.~A. Generalization of the Gaussian
  electrostatic model: Extension to arbitrary angular momentum, distributed
  multipoles, and speedup with reciprocal space methods. \emph{The Journal of
  Chemical Physics} \textbf{2006}, \emph{125}, 184101\relax
\mciteBstWouldAddEndPuncttrue
\mciteSetBstMidEndSepPunct{\mcitedefaultmidpunct}
{\mcitedefaultendpunct}{\mcitedefaultseppunct}\relax
\EndOfBibitem
\bibitem[Piquemal and Cisneros(2016)Piquemal, and Cisneros]{statusGEM}
Piquemal,~J.-P.; Cisneros,~G.~A. \emph{Status of the Gaussian Electrostatic
  Model, a Density-Based Polarizable Force Field}; Pan Standford Publishing PTE
  LTD: Singapore, 2016; pp 269--299\relax
\mciteBstWouldAddEndPuncttrue
\mciteSetBstMidEndSepPunct{\mcitedefaultmidpunct}
{\mcitedefaultendpunct}{\mcitedefaultseppunct}\relax
\EndOfBibitem
\bibitem[Chaudret \latin{et~al.}(2014)Chaudret, Gresh, Narth, Lagardère,
  Darden, Cisneros, and Piquemal]{SG1}
Chaudret,~R.; Gresh,~N.; Narth,~C.; Lagardère,~L.; Darden,~T.~A.;
  Cisneros,~G.~A.; Piquemal,~J.-P. S/G-1: An ab Initio Force-Field Blending
  Frozen Hermite Gaussian Densities and Distributed Multipoles. Proof of
  Concept and First Applications to Metal Cations. \emph{The Journal of
  Physical Chemistry A} \textbf{2014}, \emph{118}, 7598--7612, PMID:
  24878003\relax
\mciteBstWouldAddEndPuncttrue
\mciteSetBstMidEndSepPunct{\mcitedefaultmidpunct}
{\mcitedefaultendpunct}{\mcitedefaultseppunct}\relax
\EndOfBibitem
\bibitem[Misquitta \latin{et~al.}(2014)Misquitta, Stone, and Fazeli]{ISA}
Misquitta,~A.~J.; Stone,~A.~J.; Fazeli,~F. Distributed Multipoles from a Robust
  Basis-Space Implementation of the Iterated Stockholder Atoms Procedure.
  \emph{Journal of Chemical Theory and Computation} \textbf{2014}, \emph{10},
  5405--5418, PMID: 26583224\relax
\mciteBstWouldAddEndPuncttrue
\mciteSetBstMidEndSepPunct{\mcitedefaultmidpunct}
{\mcitedefaultendpunct}{\mcitedefaultseppunct}\relax
\EndOfBibitem
\bibitem[Piquemal \latin{et~al.}(2003)Piquemal, Gresh, and
  Giessner-Prettre]{Piquemal2003}
Piquemal,~J.-P.; Gresh,~N.; Giessner-Prettre,~C. Improved Formulas for the
  Calculation of the Electrostatic Contribution to the Intermolecular
  Interaction Energy from Multipolar Expansion of the Electronic Distribution.
  \emph{The Journal of Physical Chemistry A} \textbf{2003}, \emph{107},
  10353--10359, PMID: 26313624\relax
\mciteBstWouldAddEndPuncttrue
\mciteSetBstMidEndSepPunct{\mcitedefaultmidpunct}
{\mcitedefaultendpunct}{\mcitedefaultseppunct}\relax
\EndOfBibitem
\bibitem[Wang \latin{et~al.}(2015)Wang, Rackers, He, Qi, Narth, Lagard{\`e}re,
  Gresh, Ponder, Piquemal, and Ren]{doi:10.1021/acs.jctc.5b00267}
Wang,~Q.; Rackers,~J.~A.; He,~C.; Qi,~R.; Narth,~C.; Lagard{\`e}re,~L.;
  Gresh,~N.; Ponder,~J.~W.; Piquemal,~J.-P.; Ren,~P. General Model for Treating
  Short-Range Electrostatic Penetration in a Molecular Mechanics Force Field.
  \emph{Journal of Chemical Theory and Computation} \textbf{2015}, \emph{11},
  2609--2618, PMID: 26413036\relax
\mciteBstWouldAddEndPuncttrue
\mciteSetBstMidEndSepPunct{\mcitedefaultmidpunct}
{\mcitedefaultendpunct}{\mcitedefaultseppunct}\relax
\EndOfBibitem
\bibitem[Rackers \latin{et~al.}(2017)Rackers, Wang, Liu, Piquemal, Ren, and
  Ponder]{C6CP06017J}
Rackers,~J.~A.; Wang,~Q.; Liu,~C.; Piquemal,~J.-P.; Ren,~P.; Ponder,~J.~W. An
  optimized charge penetration model for use with the AMOEBA force field.
  \emph{Phys. Chem. Chem. Phys.} \textbf{2017}, \emph{19}, 276--291\relax
\mciteBstWouldAddEndPuncttrue
\mciteSetBstMidEndSepPunct{\mcitedefaultmidpunct}
{\mcitedefaultendpunct}{\mcitedefaultseppunct}\relax
\EndOfBibitem
\bibitem[Narth \latin{et~al.}(2016)Narth, Lagardère, Polack, Gresh, Wang,
  Bell, Rackers, Ponder, Ren, and Piquemal]{NarthJCC}
Narth,~C.; Lagardère,~L.; Polack,~E.; Gresh,~N.; Wang,~Q.; Bell,~D.~R.;
  Rackers,~J.~A.; Ponder,~J.~W.; Ren,~P.~Y.; Piquemal,~J.-P. Scalable
  improvement of SPME multipolar electrostatics in anisotropic polarizable
  molecular mechanics using a general short-range penetration correction up to
  quadrupoles. \emph{Journal of Computational Chemistry} \textbf{2016},
  \emph{37}, 494--506\relax
\mciteBstWouldAddEndPuncttrue
\mciteSetBstMidEndSepPunct{\mcitedefaultmidpunct}
{\mcitedefaultendpunct}{\mcitedefaultseppunct}\relax
\EndOfBibitem
\bibitem[Stone and Alderton(1985)Stone, and Alderton]{DMA}
Stone,~A.; Alderton,~M. Distributed multipole analysis. \emph{Molecular
  Physics} \textbf{1985}, \emph{56}, 1047--1064\relax
\mciteBstWouldAddEndPuncttrue
\mciteSetBstMidEndSepPunct{\mcitedefaultmidpunct}
{\mcitedefaultendpunct}{\mcitedefaultseppunct}\relax
\EndOfBibitem
\bibitem[Murrell and Teixeira-Dias(1970)Murrell, and
  Teixeira-Dias]{Murelloverlap1}
Murrell,~J.; Teixeira-Dias,~J. The dependence of exchange energy on orbital
  overlap. \emph{Molecular Physics} \textbf{1970}, \emph{19}, 521--531\relax
\mciteBstWouldAddEndPuncttrue
\mciteSetBstMidEndSepPunct{\mcitedefaultmidpunct}
{\mcitedefaultendpunct}{\mcitedefaultseppunct}\relax
\EndOfBibitem
\bibitem[Williams \latin{et~al.}(1967)Williams, Schaad, and
  Murrell]{Murrelloverlap2}
Williams,~D.~R.; Schaad,~L.~J.; Murrell,~J.~N. Deviations from Pairwise
  Additivity in Intermolecular Potentials. \emph{The Journal of Chemical
  Physics} \textbf{1967}, \emph{47}, 4916--4922\relax
\mciteBstWouldAddEndPuncttrue
\mciteSetBstMidEndSepPunct{\mcitedefaultmidpunct}
{\mcitedefaultendpunct}{\mcitedefaultseppunct}\relax
\EndOfBibitem
\bibitem[Salem and Longuet-Higgins(1961)Salem, and Longuet-Higgins]{Salem}
Salem,~L.; Longuet-Higgins,~H.~C. The forces between polyatomic molecules. II.
  Short-range repulsive forces. \emph{Proceedings of the Royal Society of
  London. Series A. Mathematical and Physical Sciences} \textbf{1961},
  \emph{264}, 379--391\relax
\mciteBstWouldAddEndPuncttrue
\mciteSetBstMidEndSepPunct{\mcitedefaultmidpunct}
{\mcitedefaultendpunct}{\mcitedefaultseppunct}\relax
\EndOfBibitem
\bibitem[Chaudret \latin{et~al.}(2011)Chaudret, Gresh, Parisel, and
  Piquemal]{robinexchange}
Chaudret,~R.; Gresh,~N.; Parisel,~O.; Piquemal,~J.-P. Many-body
  exchange-repulsion in polarizable molecular mechanics. I. orbital-based
  approximations and applications to hydrated metal cation complexes.
  \emph{Journal of Computational Chemistry} \textbf{2011}, \emph{32},
  2949--2957\relax
\mciteBstWouldAddEndPuncttrue
\mciteSetBstMidEndSepPunct{\mcitedefaultmidpunct}
{\mcitedefaultendpunct}{\mcitedefaultseppunct}\relax
\EndOfBibitem
\bibitem[Foster and Boys(1960)Foster, and Boys]{Boys}
Foster,~J.~M.; Boys,~S.~F. Canonical Configurational Interaction Procedure.
  \emph{Rev. Mod. Phys.} \textbf{1960}, \emph{32}, 300--302\relax
\mciteBstWouldAddEndPuncttrue
\mciteSetBstMidEndSepPunct{\mcitedefaultmidpunct}
{\mcitedefaultendpunct}{\mcitedefaultseppunct}\relax
\EndOfBibitem
\bibitem[El~Khoury \latin{et~al.}(2017)El~Khoury, Naseem-Khan, Kwapien,
  Hobaika, Maroun, Piquemal, and Gresh]{khourysmeared}
El~Khoury,~L.; Naseem-Khan,~S.; Kwapien,~K.; Hobaika,~Z.; Maroun,~R.~G.;
  Piquemal,~J.-P.; Gresh,~N. Importance of explicit smeared lone-pairs in
  anisotropic polarizable molecular mechanics. Torture track angular tests for
  exchange-repulsion and charge transfer contributions. \emph{Journal of
  Computational Chemistry} \textbf{2017}, \emph{38}, 1897--1920\relax
\mciteBstWouldAddEndPuncttrue
\mciteSetBstMidEndSepPunct{\mcitedefaultmidpunct}
{\mcitedefaultendpunct}{\mcitedefaultseppunct}\relax
\EndOfBibitem
\bibitem[Thole(1981)]{THOLE1981341}
Thole,~B. Molecular polarizabilities calculated with a modified dipole
  interaction. \emph{Chemical Physics} \textbf{1981}, \emph{59}, 341--350\relax
\mciteBstWouldAddEndPuncttrue
\mciteSetBstMidEndSepPunct{\mcitedefaultmidpunct}
{\mcitedefaultendpunct}{\mcitedefaultseppunct}\relax
\EndOfBibitem
\bibitem[Lipparini \latin{et~al.}(2014)Lipparini, Lagardère, Stamm, Cancès,
  Schnieders, Ren, Maday, and Piquemal]{scalablelip}
Lipparini,~F.; Lagardère,~L.; Stamm,~B.; Cancès,~E.; Schnieders,~M.; Ren,~P.;
  Maday,~Y.; Piquemal,~J.-P. Scalable Evaluation of Polarization Energy and
  Associated Forces in Polarizable Molecular Dynamics: I. Toward Massively
  Parallel Direct Space Computations. \emph{Journal of Chemical Theory and
  Computation} \textbf{2014}, \emph{10}, 1638--1651, PMID: 26512230\relax
\mciteBstWouldAddEndPuncttrue
\mciteSetBstMidEndSepPunct{\mcitedefaultmidpunct}
{\mcitedefaultendpunct}{\mcitedefaultseppunct}\relax
\EndOfBibitem
\bibitem[Lagard{\`e}re \latin{et~al.}(2015)Lagard{\`e}re, Lipparini, Polack,
  Stamm, Cances, Schnieders, Ren, Maday, and Piquemal]{lagardere2015scalable}
Lagard{\`e}re,~L.; Lipparini,~F.; Polack,~E.; Stamm,~B.; Cances,~E.;
  Schnieders,~M.; Ren,~P.; Maday,~Y.; Piquemal,~J.-P. Scalable evaluation of
  polarization energy and associated forces in polarizable molecular dynamics:
  II. Toward massively parallel computations using smooth particle mesh Ewald.
  \emph{Journal of chemical theory and computation} \textbf{2015}, \emph{11},
  2589--2599\relax
\mciteBstWouldAddEndPuncttrue
\mciteSetBstMidEndSepPunct{\mcitedefaultmidpunct}
{\mcitedefaultendpunct}{\mcitedefaultseppunct}\relax
\EndOfBibitem
\bibitem[Aviat \latin{et~al.}(2017)Aviat, Levitt, Stamm, Maday, Ren, Ponder,
  Lagard{\`{e}}re, and Piquemal]{Aviat2017}
Aviat,~F.; Levitt,~A.; Stamm,~B.; Maday,~Y.; Ren,~P.; Ponder,~J.~W.;
  Lagard{\`{e}}re,~L.; Piquemal,~J.~P. {Truncated conjugate gradient: An
  optimal strategy for the analytical evaluation of the many-body polarization
  energy and forces in molecular simulations}. \emph{J. Chem. Theory Comput.}
  \textbf{2017}, \emph{13}, 180--190\relax
\mciteBstWouldAddEndPuncttrue
\mciteSetBstMidEndSepPunct{\mcitedefaultmidpunct}
{\mcitedefaultendpunct}{\mcitedefaultseppunct}\relax
\EndOfBibitem
\bibitem[Aviat \latin{et~al.}(2017)Aviat, Lagard{\`{e}}re, and
  Piquemal]{Aviat2017a}
Aviat,~F.; Lagard{\`{e}}re,~L.; Piquemal,~J.~P. {The truncated conjugate
  gradient (TCG), a non-iterative/fixed-cost strategy for computing
  polarization in molecular dynamics: Fast evaluation of analytical forces}.
  \emph{J. Chem. Phys.} \textbf{2017}, \emph{147}, 161724\relax
\mciteBstWouldAddEndPuncttrue
\mciteSetBstMidEndSepPunct{\mcitedefaultmidpunct}
{\mcitedefaultendpunct}{\mcitedefaultseppunct}\relax
\EndOfBibitem
\bibitem[Lagardère \latin{et~al.}(2019)Lagardère, Aviat, and
  Piquemal]{pushing}
Lagardère,~L.; Aviat,~F.; Piquemal,~J.-P. Pushing the Limits of
  Multiple-Time-Step Strategies for Polarizable Point Dipole Molecular
  Dynamics. \emph{The Journal of Physical Chemistry Letters} \textbf{2019},
  \emph{10}, 2593--2599\relax
\mciteBstWouldAddEndPuncttrue
\mciteSetBstMidEndSepPunct{\mcitedefaultmidpunct}
{\mcitedefaultendpunct}{\mcitedefaultseppunct}\relax
\EndOfBibitem
\bibitem[Murrell \latin{et~al.}(1965)Murrell, Randić, Williams, and
  Longuet-Higgins]{MurellTC}
Murrell,~J.~N.; Randić,~M.; Williams,~D.~R.; Longuet-Higgins,~H.~C. The theory
  of intermolecular forces in the region of small orbital overlap.
  \emph{Proceedings of the Royal Society of London. Series A. Mathematical and
  Physical Sciences} \textbf{1965}, \emph{284}, 566--581\relax
\mciteBstWouldAddEndPuncttrue
\mciteSetBstMidEndSepPunct{\mcitedefaultmidpunct}
{\mcitedefaultendpunct}{\mcitedefaultseppunct}\relax
\EndOfBibitem
\bibitem[Misquitta(2013)]{Misquitta13}
Misquitta,~A.~J. Charge-transfer from Regularized Symmetry-Adapted Perturbation
  Theory. \emph{J. Chem. Theory Comput.} \textbf{2013}, \emph{9},
  5313--5326\relax
\mciteBstWouldAddEndPuncttrue
\mciteSetBstMidEndSepPunct{\mcitedefaultmidpunct}
{\mcitedefaultendpunct}{\mcitedefaultseppunct}\relax
\EndOfBibitem
\bibitem[hsl()]{hsl}
The HSL Mathematical Software Library; Science and Technology Facilities
  Council. \url{http://www.hsl.rl.ac.uk/}\relax
\mciteBstWouldAddEndPuncttrue
\mciteSetBstMidEndSepPunct{\mcitedefaultmidpunct}
{\mcitedefaultendpunct}{\mcitedefaultseppunct}\relax
\EndOfBibitem
\bibitem[Liu \latin{et~al.}(2020)Liu, Piquemal, and Ren]{AMOEBA+2}
Liu,~C.; Piquemal,~J.-P.; Ren,~P. Implementation of Geometry-Dependent Charge
  Flux into the Polarizable AMOEBA+ Potential. \emph{The Journal of Physical
  Chemistry Letters} \textbf{2020}, \emph{11}, 419--426, PMID: 31865706\relax
\mciteBstWouldAddEndPuncttrue
\mciteSetBstMidEndSepPunct{\mcitedefaultmidpunct}
{\mcitedefaultendpunct}{\mcitedefaultseppunct}\relax
\EndOfBibitem
\bibitem[Reinhardt and Piquemal(2009)Reinhardt, and Piquemal]{ReinhPiqwater}
Reinhardt,~P.; Piquemal,~J.-P. New intermolecular benchmark calculations on the
  water dimer: SAPT and supermolecular post-Hartree–Fock approaches.
  \emph{International Journal of Quantum Chemistry} \textbf{2009}, \emph{109},
  3259--3267\relax
\mciteBstWouldAddEndPuncttrue
\mciteSetBstMidEndSepPunct{\mcitedefaultmidpunct}
{\mcitedefaultendpunct}{\mcitedefaultseppunct}\relax
\EndOfBibitem
\bibitem[Soper and Phillips(1986)Soper, and Phillips]{Soper1986}
Soper,~A.~K.; Phillips,~M.~G. {A new determination of the structure of water at
  25$^{\circ}$C}. \emph{Chem. Phys.} \textbf{1986}, \emph{107}, 47--60\relax
\mciteBstWouldAddEndPuncttrue
\mciteSetBstMidEndSepPunct{\mcitedefaultmidpunct}
{\mcitedefaultendpunct}{\mcitedefaultseppunct}\relax
\EndOfBibitem
\bibitem[Soper(2000)]{Soper2000}
Soper,~A.~K. {The radial distribution functions of water and ice from 220 to
  673 K and at pressures up to 400 MPa}. \emph{Chem. Phys.} \textbf{2000},
  \emph{258}, 121--137\relax
\mciteBstWouldAddEndPuncttrue
\mciteSetBstMidEndSepPunct{\mcitedefaultmidpunct}
{\mcitedefaultendpunct}{\mcitedefaultseppunct}\relax
\EndOfBibitem
\bibitem[Skinner \latin{et~al.}(2013)Skinner, Huang, Schlesinger, Pettersson,
  Nilsson, and Benmore]{Skinner2013}
Skinner,~L.~B.; Huang,~C.; Schlesinger,~D.; Pettersson,~L.~G.; Nilsson,~A.;
  Benmore,~C.~J. {Benchmark oxygen-oxygen pair-distribution function of ambient
  water from x-ray diffraction measurements with a wide Q-range}. \emph{J.
  Chem. Phys.} \textbf{2013}, \emph{138}, 74506\relax
\mciteBstWouldAddEndPuncttrue
\mciteSetBstMidEndSepPunct{\mcitedefaultmidpunct}
{\mcitedefaultendpunct}{\mcitedefaultseppunct}\relax
\EndOfBibitem
\bibitem[Fanourgakis \latin{et~al.}(2006)Fanourgakis, Schenter, and
  Xantheas]{Fanourgakis2006}
Fanourgakis,~G.~S.; Schenter,~G.~K.; Xantheas,~S.~S. {A quantitative account of
  quantum effects in liquid water}. \emph{J. Chem. Phys} \textbf{2006},
  \emph{125}, 141102\relax
\mciteBstWouldAddEndPuncttrue
\mciteSetBstMidEndSepPunct{\mcitedefaultmidpunct}
{\mcitedefaultendpunct}{\mcitedefaultseppunct}\relax
\EndOfBibitem
\bibitem[Paesani \latin{et~al.}(2007)Paesani, Iuchi, and Voth]{Paesani2007}
Paesani,~F.; Iuchi,~S.; Voth,~G.~A. {Quantum effects in liquid water from an ab
  initio -based polarizable force field}. \emph{J. Chem. Phys.} \textbf{2007},
  \emph{127}, 074506\relax
\mciteBstWouldAddEndPuncttrue
\mciteSetBstMidEndSepPunct{\mcitedefaultmidpunct}
{\mcitedefaultendpunct}{\mcitedefaultseppunct}\relax
\EndOfBibitem
\bibitem[Levitt \latin{et~al.}(1997)Levitt, Hirshberg, Sharon, Laidig, and
  Daggett]{Levitt1997}
Levitt,~M.; Hirshberg,~M.; Sharon,~R.; Laidig,~K.~E.; Daggett,~V. {Calibration
  and testing of a water model for simulation of the molecular dynamics of
  proteins and nucleic acids in solution}. \emph{J. Phys. Chem. B}
  \textbf{1997}, \emph{101}, 5051--5061\relax
\mciteBstWouldAddEndPuncttrue
\mciteSetBstMidEndSepPunct{\mcitedefaultmidpunct}
{\mcitedefaultendpunct}{\mcitedefaultseppunct}\relax
\EndOfBibitem
\bibitem[Yeh and Hummer(2004)Yeh, and Hummer]{Yeh2004}
Yeh,~I.~C.; Hummer,~G. {System-size dependence of diffusion coefficients and
  viscosities from molecular dynamics simulations with periodic boundary
  conditions}. \emph{J. Phys. Chem. B} \textbf{2004}, \emph{108},
  15873--15879\relax
\mciteBstWouldAddEndPuncttrue
\mciteSetBstMidEndSepPunct{\mcitedefaultmidpunct}
{\mcitedefaultendpunct}{\mcitedefaultseppunct}\relax
\EndOfBibitem
\bibitem[Gresh(1997)]{gresh1997}
Gresh,~N. Model, Multiply Hydrogen-Bonded Water Oligomers (N= 3- 20). How
  Closely Can a Separable, ab Initio-Grounded Molecular Mechanics Procedure
  Reproduce the Results of Supermolecule Quantum Chemical Computations?
  \emph{The Journal of Physical Chemistry A} \textbf{1997}, \emph{101},
  8680--8694\relax
\mciteBstWouldAddEndPuncttrue
\mciteSetBstMidEndSepPunct{\mcitedefaultmidpunct}
{\mcitedefaultendpunct}{\mcitedefaultseppunct}\relax
\EndOfBibitem
\bibitem[Guo \latin{et~al.}(2000)Guo, Gresh, Roques, and Salahub]{guo2000}
Guo,~H.; Gresh,~N.; Roques,~B.~P.; Salahub,~D.~R. Many-body effects in systems
  of peptide hydrogen-bonded networks and their contributions to ligand
  binding: a comparison of the performances of DFT and polarizable molecular
  mechanics. \emph{The Journal of Physical Chemistry B} \textbf{2000},
  \emph{104}, 9746--9754\relax
\mciteBstWouldAddEndPuncttrue
\mciteSetBstMidEndSepPunct{\mcitedefaultmidpunct}
{\mcitedefaultendpunct}{\mcitedefaultseppunct}\relax
\EndOfBibitem
\bibitem[Antony \latin{et~al.}(2005)Antony, Piquemal, and Gresh]{antony2005}
Antony,~J.; Piquemal,~J.-P.; Gresh,~N. Complexes of thiomandelate and captopril
  mercaptocarboxylate inhibitors to metallo-β-lactamase by polarizable
  molecular mechanics. Validation on model binding sites by quantum chemistry.
  \emph{Journal of Computational Chemistry} \textbf{2005}, \emph{26},
  1131--1147\relax
\mciteBstWouldAddEndPuncttrue
\mciteSetBstMidEndSepPunct{\mcitedefaultmidpunct}
{\mcitedefaultendpunct}{\mcitedefaultseppunct}\relax
\EndOfBibitem
\bibitem[Roux \latin{et~al.}(2007)Roux, Gresh, Perera, Piquemal, and
  Salmon]{Roux2007}
Roux,~C.; Gresh,~N.; Perera,~L.~E.; Piquemal,~J.-P.; Salmon,~L. Binding of
  5-phospho-D-arabinonohydroxamate and 5-phospho-D-arabinonate inhibitors to
  zinc phosphomannose isomerase from Candida albicans studied by polarizable
  molecular mechanics and quantum mechanics. \emph{Journal of Computational
  Chemistry} \textbf{2007}, \emph{28}, 938--957\relax
\mciteBstWouldAddEndPuncttrue
\mciteSetBstMidEndSepPunct{\mcitedefaultmidpunct}
{\mcitedefaultendpunct}{\mcitedefaultseppunct}\relax
\EndOfBibitem
\bibitem[Piquemal \latin{et~al.}(2007)Piquemal, Chelli, Procacci, and
  Gresh]{piquemal2007key}
Piquemal,~J.-P.; Chelli,~R.; Procacci,~P.; Gresh,~N. Key role of the
  polarization anisotropy of water in modeling classical polarizable force
  fields. \emph{The Journal of Physical Chemistry A} \textbf{2007}, \emph{111},
  8170--8176\relax
\mciteBstWouldAddEndPuncttrue
\mciteSetBstMidEndSepPunct{\mcitedefaultmidpunct}
{\mcitedefaultendpunct}{\mcitedefaultseppunct}\relax
\EndOfBibitem
\bibitem[Gresh \latin{et~al.}(2010)Gresh, Audiffren, Piquemal, de~Ruyck,
  Ledecq, and Wouters]{bimetallic}
Gresh,~N.; Audiffren,~N.; Piquemal,~J.-P.; de~Ruyck,~J.; Ledecq,~M.;
  Wouters,~J. Analysis of the Interactions Taking Place in the Recognition Site
  of a Bimetallic Mg(II) Zn(II) Enzyme, Isopentenyl Diphosphate Isomerase. A
  Parallel Quantum-Chemical and Polarizable Molecular Mechanics Study.
  \emph{The Journal of Physical Chemistry B} \textbf{2010}, \emph{114},
  4884--4895, PMID: 20329783\relax
\mciteBstWouldAddEndPuncttrue
\mciteSetBstMidEndSepPunct{\mcitedefaultmidpunct}
{\mcitedefaultendpunct}{\mcitedefaultseppunct}\relax
\EndOfBibitem
\bibitem[Goldwaser \latin{et~al.}(2014)Goldwaser, de~Courcy, Demange, Garbay,
  Raynaud, Hadj-Slimane, Piquemal, and Gresh]{goldwaser2014conformational}
Goldwaser,~E.; de~Courcy,~B.; Demange,~L.; Garbay,~C.; Raynaud,~F.;
  Hadj-Slimane,~R.; Piquemal,~J.-P.; Gresh,~N. Conformational analysis of a
  polyconjugated protein-binding ligand by joint quantum chemistry and
  polarizable molecular mechanics. Addressing the issues of anisotropy,
  conjugation, polarization, and multipole transferability. \emph{Journal of
  molecular modeling} \textbf{2014}, \emph{20}, 1--24\relax
\mciteBstWouldAddEndPuncttrue
\mciteSetBstMidEndSepPunct{\mcitedefaultmidpunct}
{\mcitedefaultendpunct}{\mcitedefaultseppunct}\relax
\EndOfBibitem
\bibitem[Gresh \latin{et~al.}(2015)Gresh, El~Hage, Goldwaser, De~Courcy,
  Chaudret, Perahia, Narth, Lagard{\`e}re, Lipparini, and Piquemal]{addressing}
Gresh,~N.; El~Hage,~K.; Goldwaser,~E.; De~Courcy,~B.; Chaudret,~R.;
  Perahia,~D.; Narth,~C.; Lagard{\`e}re,~L.; Lipparini,~F.; Piquemal,~J.-P.
  \emph{Quantum Modeling of Complex Molecular Systems}; Springer, 2015; pp
  1--49\relax
\mciteBstWouldAddEndPuncttrue
\mciteSetBstMidEndSepPunct{\mcitedefaultmidpunct}
{\mcitedefaultendpunct}{\mcitedefaultseppunct}\relax
\EndOfBibitem
\bibitem[Devillers \latin{et~al.}(2020)Devillers, Piquemal, Salmon, and
  Gresh]{Devillers2020}
Devillers,~M.; Piquemal,~J.; Salmon,~L.; Gresh,~N. {Calibration of the
  dianionic phosphate group: Validation on the recognition site of the
  homodimeric enzyme phosphoglucose isomerase}. \emph{J. Comput. Chem.}
  \textbf{2020}, \emph{41}, 839--854\relax
\mciteBstWouldAddEndPuncttrue
\mciteSetBstMidEndSepPunct{\mcitedefaultmidpunct}
{\mcitedefaultendpunct}{\mcitedefaultseppunct}\relax
\EndOfBibitem
\bibitem[El~Darazi \latin{et~al.}(2020)El~Darazi, El~Khoury, El~Hage, Maroun,
  Hobaika, Piquemal, and Gresh]{darazi2020}
El~Darazi,~P.; El~Khoury,~L.; El~Hage,~K.; Maroun,~R.~G.; Hobaika,~Z.;
  Piquemal,~J.-P.; Gresh,~N. Quantum-Chemistry Based Design of Halobenzene
  Derivatives With Augmented Affinities for the HIV-1 Viral G4/C16 Base-Pair.
  \emph{Frontiers in Chemistry} \textbf{2020}, \emph{8}, 440\relax
\mciteBstWouldAddEndPuncttrue
\mciteSetBstMidEndSepPunct{\mcitedefaultmidpunct}
{\mcitedefaultendpunct}{\mcitedefaultseppunct}\relax
\EndOfBibitem
\bibitem[Poier \latin{et~al.}(2022)Poier, Lagardère, and
  Piquemal]{poier_lagardere_piquemal_2021}
Poier,~P.~P.; Lagardère,~L.; Piquemal,~J.-P. O(N) Stochastic Evaluation of
  Many-Body van der Waals Energies in Large Complex Systems. \emph{Journal of
  Chemical Theory and Computation} \textbf{2022}, \emph{18}, 1633--1645, PMID:
  35133157\relax
\mciteBstWouldAddEndPuncttrue
\mciteSetBstMidEndSepPunct{\mcitedefaultmidpunct}
{\mcitedefaultendpunct}{\mcitedefaultseppunct}\relax
\EndOfBibitem
\bibitem[Mauger \latin{et~al.}(2021)Mauger, Pl{\'{e}}, Lagard{\`{e}}re,
  Bonella, Mangaud, Piquemal, and Huppert]{mauger2021}
Mauger,~N.; Pl{\'{e}},~T.; Lagard{\`{e}}re,~L.; Bonella,~S.; Mangaud,~{\'{E}}.;
  Piquemal,~J.-P.; Huppert,~S. Nuclear Quantum Effects in Liquid Water at Near
  Classical Computational Cost Using the Adaptive Quantum Thermal Bath.
  \emph{The Journal of Physical Chemistry Letters} \textbf{2021}, \emph{12},
  8285--8291\relax
\mciteBstWouldAddEndPuncttrue
\mciteSetBstMidEndSepPunct{\mcitedefaultmidpunct}
{\mcitedefaultendpunct}{\mcitedefaultseppunct}\relax
\EndOfBibitem
\bibitem[Adjoua \latin{et~al.}(2021)Adjoua, Lagardère, Jolly, Durocher, Very,
  Dupays, Wang, Inizan, Célerse, Ren, Ponder, and Piquemal]{GPU}
Adjoua,~O.; Lagardère,~L.; Jolly,~L.-H.; Durocher,~A.; Very,~T.; Dupays,~I.;
  Wang,~Z.; Inizan,~T.~J.; Célerse,~F.; Ren,~P.; Ponder,~J.~W.;
  Piquemal,~J.-P. Tinker-HP: Accelerating Molecular Dynamics Simulations of
  Large Complex Systems with Advanced Point Dipole Polarizable Force Fields
  Using GPUs and Multi-GPU Systems. \emph{Journal of Chemical Theory and
  Computation} \textbf{2021}, \emph{17}, 2034--2053, PMID: 33755446\relax
\mciteBstWouldAddEndPuncttrue
\mciteSetBstMidEndSepPunct{\mcitedefaultmidpunct}
{\mcitedefaultendpunct}{\mcitedefaultseppunct}\relax
\EndOfBibitem
\bibitem[THP()]{THPGithub}
Tinker-HP: High-Performance Massively Parallel Evolution of Tinker on CPUs \&
  GPUs. \url{https://github.com/TinkerTools/tinker-hp/tree/sibfa/sibfa}\relax
\mciteBstWouldAddEndPuncttrue
\mciteSetBstMidEndSepPunct{\mcitedefaultmidpunct}
{\mcitedefaultendpunct}{\mcitedefaultseppunct}\relax
\EndOfBibitem
\bibitem[Frisch \latin{et~al.}(2009)Frisch, Trucks, Schlegel, Scuseria, Robb,
  Cheeseman, Scalmani, Barone, Mennucci, and Petersson]{frisch2009g09}
Frisch,~M.; Trucks,~G.; Schlegel,~H.; Scuseria,~G.; Robb,~M.; Cheeseman,~J.;
  Scalmani,~G.; Barone,~V.; Mennucci,~B.; Petersson,~G. G09 Gaussian Inc.
  2009\relax
\mciteBstWouldAddEndPuncttrue
\mciteSetBstMidEndSepPunct{\mcitedefaultmidpunct}
{\mcitedefaultendpunct}{\mcitedefaultseppunct}\relax
\EndOfBibitem
\bibitem[Misquitta and Stone(2021)Misquitta, and Stone]{misquitta_camcasp}
Misquitta,~A.; Stone,~A. CamCASP: a program for studying intermolecular
  interactions and for the calculation of molecular properties in distributed
  form. \emph{University of Cambridge} \textbf{2021}, \relax
\mciteBstWouldAddEndPunctfalse
\mciteSetBstMidEndSepPunct{\mcitedefaultmidpunct}
{}{\mcitedefaultseppunct}\relax
\EndOfBibitem
\bibitem[Smith \latin{et~al.}(1990)Smith, Swanton, Pople, Schaefer, and
  Radom]{Smith1990}
Smith,~B.~J.; Swanton,~D.~J.; Pople,~J.~A.; Schaefer,~H.~F.; Radom,~L.
  {Transition structures for the interchange of hydrogen atoms within the water
  dimer}. \emph{J. Chem. Phys.} \textbf{1990}, \emph{92}, 1240--1247\relax
\mciteBstWouldAddEndPuncttrue
\mciteSetBstMidEndSepPunct{\mcitedefaultmidpunct}
{\mcitedefaultendpunct}{\mcitedefaultseppunct}\relax
\EndOfBibitem
\bibitem[Dahlke \latin{et~al.}(2008)Dahlke, Olson, Leverentz, and
  Truhlar]{ErinE.Dahlke2008}
Dahlke,~E.~E.; Olson,~R.~M.; Leverentz,~H.~R.; Truhlar,~D.~G. {Assessment of
  the Accuracy of Density Functionals for Prediction of Relative Energies and
  Geometries of Low-Lying Isomers of Water Hexamers}. \textbf{2008}, \relax
\mciteBstWouldAddEndPunctfalse
\mciteSetBstMidEndSepPunct{\mcitedefaultmidpunct}
{}{\mcitedefaultseppunct}\relax
\EndOfBibitem
\bibitem[Bates and Tschumper(2009)Bates, and Tschumper]{Bates2009}
Bates,~D.~M.; Tschumper,~G.~S. {CCSD(T) complete basis set limit relative
  energies for low-lying water hexamer structures}. \emph{J. Phys. Chem. A}
  \textbf{2009}, \emph{113}, 3555--3559\relax
\mciteBstWouldAddEndPuncttrue
\mciteSetBstMidEndSepPunct{\mcitedefaultmidpunct}
{\mcitedefaultendpunct}{\mcitedefaultseppunct}\relax
\EndOfBibitem
\bibitem[Wang \latin{et~al.}(2013)Wang, Head-Gordon, Ponder, Ren, Chodera,
  Eastman, Martinez, and Pande]{Wang2013}
Wang,~L.-P.; Head-Gordon,~T.; Ponder,~J.~W.; Ren,~P.; Chodera,~J.~D.;
  Eastman,~P.~K.; Martinez,~T.~J.; Pande,~V.~S. {Systematic Improvement of a
  Classical Molecular Model of Water}. \emph{J. Phys. Chem. B} \textbf{2013},
  \emph{117}, 9956--9972\relax
\mciteBstWouldAddEndPuncttrue
\mciteSetBstMidEndSepPunct{\mcitedefaultmidpunct}
{\mcitedefaultendpunct}{\mcitedefaultseppunct}\relax
\EndOfBibitem
\bibitem[Xantheas and Apra(2004)Xantheas, and Apra]{Xantheas2004}
Xantheas,~S.~S.; Apra,~E. {The binding energies of the D-2d and S-4 water
  octamer isomers: High-level electronic structure and empirical potential
  results}. \emph{J. Chem. Phys.} \textbf{2004}, \emph{120}, 823--828\relax
\mciteBstWouldAddEndPuncttrue
\mciteSetBstMidEndSepPunct{\mcitedefaultmidpunct}
{\mcitedefaultendpunct}{\mcitedefaultseppunct}\relax
\EndOfBibitem
\bibitem[Fanourgakis \latin{et~al.}(2004)Fanourgakis, Apr{\`{a}}, and
  Xantheas]{Fanourgakis2004}
Fanourgakis,~G.~S.; Apr{\`{a}},~E.; Xantheas,~S.~S. {High-level ab initio
  calculations for the four low-lying families of minima of (H2O)20. I.
  Estimates of MP2/CBS binding energies and comparison with empirical
  potentials}. \emph{J. Chem. Phys.} \textbf{2004}, \emph{121},
  2655--2663\relax
\mciteBstWouldAddEndPuncttrue
\mciteSetBstMidEndSepPunct{\mcitedefaultmidpunct}
{\mcitedefaultendpunct}{\mcitedefaultseppunct}\relax
\EndOfBibitem
\bibitem[Bulusu \latin{et~al.}(2006)Bulusu, Yoo, Apr{\`{a}}, Xantheas, and
  Zeng]{Bulusu2006}
Bulusu,~S.; Yoo,~S.; Apr{\`{a}},~E.; Xantheas,~S.; Zeng,~X.~C. {Lowest-energy
  structures of water clusters (H 2O)n and (H 2O) 13}. \emph{J. Phys. Chem. A}
  \textbf{2006}, \emph{110}, 11781--11784\relax
\mciteBstWouldAddEndPuncttrue
\mciteSetBstMidEndSepPunct{\mcitedefaultmidpunct}
{\mcitedefaultendpunct}{\mcitedefaultseppunct}\relax
\EndOfBibitem
\bibitem[Yoo \latin{et~al.}(2010)Yoo, Apr{\`{a}}, Zeng, and Xantheas]{Yoo2010}
Yoo,~S.; Apr{\`{a}},~E.; Zeng,~X.~C.; Xantheas,~S.~S. {High-level Ab initio
  electronic structure calculations of water clusters (H2O)16 and (H2O)17: A
  new global minimum for (H2O)16}. \emph{J. Phys. Chem. Lett.} \textbf{2010},
  \emph{1}, 3122--3127\relax
\mciteBstWouldAddEndPuncttrue
\mciteSetBstMidEndSepPunct{\mcitedefaultmidpunct}
{\mcitedefaultendpunct}{\mcitedefaultseppunct}\relax
\EndOfBibitem
\end{mcitethebibliography}

\end{document}


\section{SIBFA's Intermolecular Potential: Analytical Expressions of Potential Energy Terms}
\label{formulae}

\subsection{S1: Formulation of Exchange-Repulsion}
As already described in (ref Robin), the repulsion model of SIBFA is overlap based and involves interactions between bonds and bonds, lone pairs and lone pairs and cross interactions between bonds and lone pairs.
\[E_{rep}=E_{rep-bond-bond}+E_{rep,lp-lp}+E_{rep,bond-lp}\]
wich can be decomposed in pairwise interactions such that for example:
\[E_{rep,bond-bond}=\sum_{i,j=1,i < j}^{nbonds}E_{i,j,rep-bond-bond}\]
For each of these pairs, one has to build the overlap between the orbitals of the involved bonds which can be expressed as a combination of overlap between $s$ and $p$ orbitals as written in (ref ect1). Let us denote by $a$ and $b$ the atoms forming the $i$ bond and by $c$ and $d$ the ones forming the $j$ bond. Each atom of each bond bearing a $s$ and a $p$ orbital, the overlap between the two bonds can be decomposed in 4 terms involving atomic pairs, each of them in term being made of 4 terms involving s and p orbitals, such that the overlaps are:
\[O_{ac}=O_{s_a s_c}+O_{s_a p_c}+O_{s_c p_a}+O_{p_a p_c}\]
\[O_{ad}=O_{s_a s_d}+O_{s_a p_d}+O_{s_d p_a}+O_{p_a p_d}\]
\[O_{bc}=O_{s_b s_c}+O_{s_b p_c}+O_{s_c p_b}+O_{p_b p_c}\]
\[O_{bd}=O_{s_b s_d}+O_{s_b p_d}+O_{s_d p_b}+O_{p_b p_d}\]
The term involving $s$ orbitals consist in the product of hybridation coefficents such that for example:
\[O_{s_a s_c}=C_{s_a} C_{s_c}\]
The terms involving $s$ and $p$ orbitals read:
\[O_{s_a p_c}=C_{s_a}C_{p_c}f_{ac}cos(\vv{AB},\vv{AC})\]
\[O_{s_c p_a}=C_{s_c}C_{p_a}f_{ca}cos(\vv{CD},\vv{CA})\]
with $f_{ab}$ and $f_{ca}$ tabulated values to approximate the corresponding integral, and the ones involving both $p$ orbitals read, for example:
\[O_{p_a p_c}=C_{p_a}C_{p_c}2cos(\vv{AB},\vv{AC})(\vv{CD},\vv{CA})\]

These overlap terms are to be multiplied by exponential prefactors:
\[S_{ac}=M_{ac}e^{-\alpha \rho_{ac}}\]
\[S_{ac_2}=M_{ac}e^{-\alpha_2 \rho_{ac}}\]
with $\alpha$ and $\alpha_2$ two constants and 
\[M_{ac}^2=g_a g_c (1-\frac{Q_a}{N_a^{val}})(1-\frac{Q_c}{N_c^{val}})\]
with $g_a$ and $g_c$ parameters, $Q_a$ and $Q_c$ the partial charges of atoms $a$ and $c$, $N_a^{val}$ and $N_b^{val}$ the number of valence electrons of atoms $a$ and $b$. The reduced distance $\rho_{ac}$ is defined as:
\[\rho_{ac}=\frac{r_{ac}}{4\sqrt{r_{vdw_a}r_{vdw_c}}}\]
with $r_{vdw_a}$ and $r_{vdw_b}$ distances specific to atoms $a$ and $b$. Similarly, we can define $S_{ad}$, $S_{ad_2}$, $S_{bc}$, $S_{bc_2}$, $S_{bd}$, $S_{bd_2}$.
Finally let us call:
\[O_{1_{i,j}}=O_{ac}S_{ac}+O_{ad}S_{ad}+O_{bc}S_{bc}+O_{bd}S_{bd}\]
\[O_{2_{i,j}}=O_{ac}S_{ac_2}+O_{ad}S_{ad_2}+O_{bc}S_{bc_2}+O_{bd}S_{bd_2}\]
Then if we define call $\vv{r_{mid_{ab}}}$ the position vector of the midpoint of $a$ and $b$ and $\vv{r_{mid_{cd}}}$ the position vector of the midpoint of $c$ and $d$, if we write
\[\vv{r_{mid}}=\vv{r_{mid_{ab}}}-\vv{r_{mid_{cd}}}\]
we have
\[E_{i,j,rep-bond-bond}=Occ_{ab}Occ_{bd}(C_1\frac{O_{1_{i,j}}^2}{r_{mid}}+C_2\frac{O_{2_{i,j}}^2}{r_{mid}^2})\]
with $C_1$ and $C_2$ constants and $Occ_{ab}$ and $Occ_{bd}$ the bond occupation number of bond $i$ and $j$

Similarly, we have:
\[E_{rep,lp-lp}=\sum_{i,j=1,i < j}^{nlp}E_{i,j,rep-lp-lp}\]
we can define an overlap term as:
\[O_{ij}=O_{s_i s_j}+O_{s_i p_j}+O_{s_j p_i}+O_{p_i p_j}\]
Let us denote by $a$ the atom carrying the lone pair $i$ and $b$ the atom carrying the lone pair $j$. As for the bond-bond part, the term involving s orbitals consists in the product of hybridation coefficients (of the atom carrying the lone pairs), such that:
\[O_{s_i s_j}=C_{s_a}C_{s_b}\]
The terms involving $s$ and $p$ orbitals read:
\[O_{s_i p_j}=C_{s_i}C_{p_j}f_{ab}cos(\vv{BJ},\vv{BI})\]
\[O_{s_j p_i}=C_{s_j}C_{p_i}f_{ab}cos(\vv{AI},\vv{AJ})\]
The term involving both $p$ orbitals read:
\[O_{p_i p_j}=C_{p_a}C_{p_b}2cos(\vv{AI},\vv{AJ})(\vv{BJ},\vv{BI})\]
We also have exponential prefactors:
\[S_{ij}=M_{ij}e^{-\alpha\rho_{ij}}\]
\[S_{ij_2}=M_{ij}e^{-\alpha_2\rho_{ij}}\]
with
\[M_{ij}^2=g_a g_b (1-\frac{Q_i}{N_a^{val}})(1-\frac{Q_j}{N_b^{val}})\] 
with $Q_i$ and $Q_j$ the charges of both lone pairs, and:
\[\rho_{ij}=\frac{r_{ab}}{4\sqrt{r_{vdw_i}r_{vdw_j}}}\]
\[O_{1_{i,j}}=O_{i,j}S_{ij}\]
\[O_{2_{i,j}}=O_{i,j}S_{ij_2}\]
so that finally:
\[E_{i,j,rep-lp-lp}=Q_{i}Q_{j}(C_1\frac{O_{1_{i,j}}^2}{r_{ij}}+C_2\frac{O_{2_{i,j}}^2}{r_{ij}^2})\]
The cross bond-lone pair is defined in a similar manner:
\[E_{rep,bond-lp}=\sum_{i=1}^{nbond}\sum_{j=1}^{nlp}E_{i,j,bond-lp}\]
Let us denote by $a$ and $b$ the atoms forming the bond $i$ and by $c$ the atom carrying the lone pair $j$. We then have the two overlap terms:
\[O_{aj}=O_{s_a s_j}+O_{s_a p_j}+O_{s_j p_a}+O_{p_a p_j}\]
\[O_{bj}=O_{s_b s_j}+O_{s_b p_j}+O_{s_j p_b}+O_{p_b p_j}\]
Then, we have for example:
\[O_{s_a s_j}=C_{s_a}C_{s_j}\]
\[O_{s_a p_j}=C_{s_a}C_{p_j}f_{ac}cos(\vv{CA},\vv{CI})\]
\[O_{p_a p_j}=C_{p_a}C_{p_j}f_{ac}cos(\vv{AB},\vv{AC})cos(\vv{CA},\vv{CI})\]

\[S_{aj}=M_{aj}e^{-\alpha\rho_{aj}}\]
\[S_{aj_2}=M_{aj}e^{-\alpha_2\rho_{aj}}\] with
\[M_{aj}^2=g_a g_k (1-\frac{Q_a}{N_a^{val}})(1-\frac{Q_c}{N_c^{val}})\] 
\[\rho_{aj}=\frac{r_{ak}}{4\sqrt{r_{vdw_a}r_{vdw_j}}}\]
\[O_{1_{i,j}}=O_{aj}S_{aj}+O_{bj}S_{bj}\]
\[O_{2_{i,j}}=O_{aj}S_{aj_2}+O_{bj}S_{bj_2}\]
Then if we define call $\vv{r_{mid_{ab}}}$ the position vector of the midpoint of $a$ and $b$ and by $\vv{r}=\vv{r_{mid_{ab}}}-\vv{r_j}$
\[E_{i,j,rep-bond-lp}=Occ_{ab}Q_{j}(C_1\frac{O_{1_{i,j}}^2}{r}+C_2\frac{O_{2_{i,j}}^2}{r^2})\]

\subsection{Polarization}
The polarization energy used here is similar to the one used in AMOEBA: we use isotropic scalar polarizabilites and Thole damping at short range but with variable damping factor contrary to what is done in AMOEBA. Computationally, the induced dipoles are converged using the same machinery as is commonly used with AMOEBA and Tinker-HP: preconditioned conjugate gradient.\cite{Aviat2017,Aviat2017a}
\subsection{Many body Charge Transfer}
MRW original formulae and where it comes from, simplification for the denominator that reduces complexity
Charge transfer between an electron donnor molecule A and an electron acceptor molecule B.
\[E_{ct}=-2\sum_{\alpha}\sum_{\beta^*}\frac{I_{\alpha\beta^*}^2}{\Delta E_{\alpha\beta^*}}\]
where $\alpha$ are the occupied molecular orbitals of A and $\beta^*$ are the unoccupied molecular orbitals of B. $\Delta E_{\alpha\beta^*}$ is the energy associated to the electron transfer between $\beta^*$ and $\alpha$ and
\[I_{\alpha\beta^*}=\int \rho_{\alpha\beta^*}(r)V(r)dv\]
and where $\rho_{\alpha\beta^*}$ is the overlap transition density:
\[\rho_{\alpha\beta^*}=-(\alpha\beta^*-\alpha^2S_{\alpha\beta^*})\]
where $S_{\alpha\beta^*}$ is the overlap integral between $\alpha$ and $\beta^*$. A series of further approximation are described in ref (ECT1). In SIBFA, the electron donors are the lone-pairs of the molecules and the electron acceptors are the bond involving a heavy atom. For each lone pair $i_{lp}$ let us denote by $c(i_{lp})$ the index of its carrier. Let us also denote by $V_{i_{lp}}$ the electrostatic potential on $i_{lp}$ due to the complete system. For a given acceptor $k$ let us denote by $a_1(k)$ and $a_2(k)$ the index of both the atoms constituting it, $a_1(k)$ being the heavy one and $a_2(k)$ being the hydrogen one. We call $V_{a_1(k)}$ and $V_{a_2(k)}$ the electrostatics potential on $a_1(k)$ and $a_2(k)$ due to the total system except the local molecule which both atoms belong to. We approximate the intramolecular electrostatic potential on $a_1(k)$ and $a_2(k)$, $V_{int}(a_1(k))$ and $V_{int}(a_2(k))$ by using the expression of equation 24 and 25 of reference (ECT1). We denote by $Q_{i_{lp}}$ the charge of the lone pair $i_{lp}$ and we assign vdw radii to the lone pair carrier $r_{vdw}i$ and the two atoms constituting the acceptor: $r_{vdw}a_1(k)$, $r_{vdw}a_1(k)$. We then define the two reduced distances:
\[\rho_{1}=\frac{r_{c(i_{lp})a_1(k)}}{2\sqrt{r_{vdw}ir_{vdw}a_1(k)}}\]
\[\rho_{1}=\frac{r_{c(i_{lp})a_2(k)}}{2\sqrt{r_{vdw}ir_{vdw}a_2(k)}}\]
where $r_{c(i_{lp})a_1(k)}$ is the distance between the lone pair carrier and the first atom of the acceptor, and $r_{c(i_{lp})a_2(k)}$ the distance between the lone pair carrier and the second atom of the acceptor.

For each donnor-acceptor pair we define the following exponential terms:
\[e_1=e^{-\eta \rho_1}(V_{int}(a_1(k))-V_{i_{lp}})(C_st_{i_{lp}a_1(k)}+C_pm_{i_{lp}a_1(k)}cos(\alpha))\]
where $\eta$ is a parameter, $C_s$ and $C_p$ hybridization coefficient, $t_{i_{lp}a_1(k)}$ and $m_{i_{lp}a_1(k)}$ tabulated coefficient used to approximate integrals described in ref 24 and 25 and $\alpha$ the angle formed by the two vectors: $r_{c(i_{lp})i_{lp}}$ and $r_{c(i_{lp})a_1(k)}$.
\[e_2=e^{-\eta \rho_2}(V_{int}(a_2(k))-V_{i_{lp}})(C_st_{i_{lp}a_2(k)}+C_pm_{i_{lp}a_2(k)}cos(\beta))\]
with the same tabulated coefficients as for the $e_1$ term and with $\beta$ the angle formed by the two vectors: $r_{c(i_{lp})i_{lp}}$ and $r_{c(i_{lp})a_2(k)}$.
The final charge transfer energy used in SIBFA for water then reads:
\[E_{ct}=\sum_{i_{lp}}\sum_{k}-q(i_{lp})\frac{(e_1-e_2)^2}{\Delta E}\]
with 
\[\Delta E=ah(i)+V_{i_{lp}}-(ae((a_1(k))+V_{a_1(k)})\]
where $ah(i)$ is the electronic affinity of the lone pair and $ae(a_1(k))$ the ionization potential of $a_1(k)$.
\subsection{Dispersion and Exchange Dispersion}
The dispersion model used in SIBFA involves atoms and lone pairs, more precisely it contains an atom-atom term, a lone-pair-lone-pair term and a cross atom-lone pair term.
\[E_{disp}=E_{disp-at-at}+E_{disp-lp-lp}+E_{disp-at-lp}\]
each of these terms corresponding to pairwise interactions, such that:
\[E_{disp-at-at}=\sum_{i,j=1,i < j}^{natoms}E_{i,j,at-at}\]
Let us define the reduced distance:
\[m_{ij}=\frac{r_{ij}}{2\sqrt{r_{vdw_i}r_{vdw_j}}}\]
with $r_{vdw_i}$ and $r_{vdw_j}$ van der Waals radii associated to atoms i and j. Let us also define the ratio:
\[d_{ij}=\frac{(r_{vdw_i}+r_{vdw_j})\alpha}{r_{ij}}-1\]
$\alpha$ being a global parameter and the pairwise parameter:
\[\beta_{ij}=\beta_i\beta_j\]
$\beta_i$ and $\beta_j$ being constants specific to atoms $i$ and $j$, then:
\[E_{i,j,at-at}=\beta_{ij}(C_6\frac{e^{-\gamma_6 d_{ij}}}{m_{ij}^6}+C_8\frac{e^{-\gamma_8 d_{ij}}}{m_{ij}^8}+C_{10}\frac{e^{-\gamma_{10} d_{ij}}}{m_{ij}^{10}})\]
with $\gamma_6$, $\gamma_8$ and $\gamma_{10}$, $C_6$, $C_8$ and $C_{10}$ global parameters.
Similarly, we have:
\[E_{disp-lp-lp}=\sum_{k,l=1,k < l}^{nlp}E_{k,l,lp-lp}\]
\[m_{kl}=\frac{r_{kl}}{2\sqrt{r_{vdw_k}r_{vdw_l}}}\]
with $r_{vdw_k}$ and $r_{vdw_l}$ van der Waals radii associated to lone pairs k and l
\[d_{kl}=\frac{(r_{vdw_k}+r_{vdw_l})\alpha}{r_{kl}}-1\]
Then, denoting by $Q_k$ and $Q_l$ the charge of respectively lone pair $k$ and lone pair $l$:
\[E_{k,l,lp-lp}=-0.5Q_k0.5Q_l(C_6\frac{e^{-\gamma_6 d_{kl}}}{m_{kl}^6}+C_8\frac{e^{-\gamma_8 d_{kl}}}{m_{kl}^8}+C_{10}\frac{e^{-\gamma_{10} d_{kl}}}{m_{kl}^{10}})\]
Finally, the cross atom-lone pair term reads:
\[E_{disp-atom-lp}=\sum_{i=1}^{natoms}\sum_{k=1}^{nlp}E_{i,k,atom-lp}\]
Let us denote by $c(k)$ the index of the atom carrying the lone pair k.
\[m_{ik}=\frac{r_{ik}}{2\sqrt{r_{vdw_i}r_{vdw_k}}}\]
\[d_{ik}=\frac{(r_{vdw_i}+r_{vdw_k})\alpha}{r_{ik}}-1\]
\[E_{i,k,atom-lp}=-0.5\delta(C_6\frac{e^{-\gamma_6 d_{ik}}}{m_{ik}^6}+C_8\frac{e^{-\gamma_8 d_{ik}}}{m_{ik}^8}+C_{10}\frac{e^{-\gamma_{10} d_{ik}}}{m_{ik}^{10}})\]
$\delta$ being an additional scaling factor specific to atoms-lone pair dispersion.

These 3 types of pairwise interactions are complemented by exponential-like short range exchange dispersion terms, such that we can define:
\[E_{exchange-disp}=E_{xdisp-at-at}+E_{xdisp-lp-lp}+E_{xdisp-at-lp}\]
each of these terms also corresponding to pairwise interactions, such that:
\[E_{xdisp-at-at}=\sum_{i,j=1,i < j}^{natoms}Ex_{i,j,at-at}\]
Let us note $Q_i$ and $Q_j$ the partial charge carried by atom $i$ and $j$, $N_i^{val}$, and $N_j^{val}$ their number of valence electrons. 
Let us define:
\[K_{ij}=\beta_{ij}(1-\frac{Q_i}{N_i^{val}})(1-\frac{Q_j}{N_j^{val}})\]
$\beta_{ij}$ being define before. Then,
\[Ex_{i,j,at-at}=C_1 K_{ij}e^{-A_1 m_{ij}}\]
with $m_{ij}$ defined as before and $C_1$, $A_1$ parameters specific to atom-atom exchange dispersion. Similarly:
\[E_{xdisp-lp-lp}=\sum_{k,l=1,k < l}^{nlp}Ex_{k,l,lp-lp}\]
\[K_{kl}=\beta_{c(k)c(l)}(1-\frac{Q_{c(k)}}{N_{c(k)}^{val}})(1-\frac{Q_{c(l)}}{N_{c(l)}^{val}})\]
\[Ex_{k,l,lp-lp}=0.5Q_k 0.5Q_lC_2 K_{kl}e^{-A_2 m_{kl}}\]
with $m_{kl}$ defined as before and $C_2$, $A_2$ parameters specific to lone-pair lone-pair exchange dispersion.
\[E_{xdisp-at-lp}=\sum_{i=1}^{natoms}\sum_{k=1}^{nlp}Ex_{i,k,at-lp}\]
\[K_{ik}=\beta_{ic(k)}(1-\frac{Q_i}{N_i^{val}})(1-\frac{Q_{c(k)}}{N_{c(k)}^{val}})\]
\[Ex_{i,k,at-lp}=0.5Q_kC_3 K_{ik}e^{-A_3 m_{ik}}\]
with $m_{ik}$ defined as before and $C_3$, $A_3$ parameters specific to atom lone-pair exchange dispersion.



\section{Strategy to compute gradients and associated computational complexity}
\subsection{Charge Transfer}

\subsection{Dispersion}
\subsection{Repulsion}
\subsection{Electrostatics}
\subsection{Polarization}


\bibliography{biblio.bib}